\documentclass[journal]{IEEEtran}
\usepackage[table,xcdraw]{xcolor}
\usepackage{adjustbox}

\usepackage{amssymb}
\usepackage{enumerate}
\usepackage{mathtools}
\usepackage{comment}
\usepackage{supertabular}
\usepackage{amsmath}
\usepackage{caption}
\usepackage{placeins}
\usepackage{subcaption}
\usepackage{booktabs}
\usepackage{color}
\usepackage{physics}
\usepackage{amsmath}
\usepackage{algorithm}
\usepackage{algpseudocode}
\usepackage{cite}
\usepackage{color,soul}
\usepackage[colorinlistoftodos]{todonotes}
\usepackage[table,xcdraw]{xcolor}
\usepackage{gensymb}
\usepackage{flushend} 
\usepackage{multicol}
\usepackage{multirow} 
\usepackage{graphicx}
\usepackage{mwe}
\graphicspath{ {./Images/} }

\usepackage[utf8]{inputenc}
\usepackage[T1]{fontenc}
\usepackage{wrapfig}

\usepackage{cite}
\usepackage[utf8]{inputenc}
\usepackage{fancyhdr} 
\usepackage{tikz}
\usetikzlibrary{shapes,backgrounds}
\usepackage{verbatim}
\usepackage{url}
\usepackage{enumitem}
\usepackage{dirtytalk}
\usepackage{amsmath,amssymb,amsfonts}
\usepackage{xfrac}
\usepackage{graphicx}
\usepackage{textcomp}
\usepackage{xcolor}
\usepackage{color,soul}
\usepackage{scalerel}
\usepackage{csquotes}
\usepackage{subcaption}
\usepackage{multirow}
\usepackage{tikz}
\usetikzlibrary{svg.path}
\usepackage{nomencl}
\makenomenclature

\usetikzlibrary{svg.path}

\definecolor{googleblue}{rgb}{0.259,0.522,0.957}
\definecolor{googlegreen}{rgb}{0.060,0.620,0.350}
\definecolor{googlered}{rgb}{0.859,0.267,0.220}
\definecolor{bleudefrance}{rgb}{0.19,0.55,0.91}

\definecolor{orcidlogocol}{HTML}{A6CE39}
\tikzset{
  orcidlogo/.pic={
    \fill[orcidlogocol] svg{M256,128c0,70.7-57.3,128-128,128C57.3,256,0,198.7,0,128C0,57.3,57.3,0,128,0C198.7,0,256,57.3,256,128z};
    \fill[white] svg{M86.3,186.2H70.9V79.1h15.4v48.4V186.2z}
                 svg{M108.9,79.1h41.6c39.6,0,57,28.3,57,53.6c0,27.5-21.5,53.6-56.8,53.6h-41.8V79.1z M124.3,172.4h24.5c34.9,0,42.9-26.5,42.9-39.7c0-21.5-13.7-39.7-43.7-39.7h-23.7V172.4z}
                 svg{M88.7,56.8c0,5.5-4.5,10.1-10.1,10.1c-5.6,0-10.1-4.6-10.1-10.1c0-5.6,4.5-10.1,10.1-10.1C84.2,46.7,88.7,51.3,88.7,56.8z};
  }
}

\newcommand\orcidicon[1]{\href{https://orcid.org/#1}{\mbox{\scalerel*{
\begin{tikzpicture}[yscale=-1,transform shape]
\pic{orcidlogo};
\end{tikzpicture}
}{|}}}}

\usepackage[bookmarks=false]{hyperref} 
\hypersetup{
    colorlinks=true,
}

\begin{document}
\raggedbottom

\twocolumn
\setcounter{page}{1}

\title{Enhanced Thermal Management in High-Temperature Applications: Design and Optimization of a Water-Cooled Forced Convection System in a Hollow Cuboid Vapour Chamber Using COMSOL and MATLAB}

\author{
        \IEEEauthorblockN{Brandon C. Colelough,~$^{\orcidicon{0000-0001-8389-3403}}$}
        
        \IEEEauthorblockA{School of Engineering and Information Technology, University of New South Wales, Australia}
}

\maketitle

\begin{abstract}

This report details the design and optimisation of a water-cooled forced convection heat dissipation system for use in high-temperature applications (ranges between $700\degree$ - $1000 \degree$K). A hollow cuboid vapour chamber model was investigated. The space within the hollow cuboid was used as the design space. COMSOL, a FEM software product was used to solve for the physical parameters of each geometry for the heat dissipation system design space. COMSOL in conjunction with MATLAB was used for the parametric and density-based topology optimisation of the geometric design in the design space. The goal of the optimization is the minimisation of a temperature gradient over the design space. This allows the heat to be evenly spread throughout the designed mesh which allows for more effective cooling. To reduce the computational time needed to solve and optimise each geometry in 3D, a 2D representation was created for the front and rear faces of the hollow cuboid setup. These 2D face designs were then extrapolated into 3D over the length of the hollow cube and COMSOL was used to find a solution for each model. This report also proposes a use case for this system wherein it would be used in conjunction with MGA and thermometric technology within coal-fired power stations for the extraction and storage of waste heat for later use.   

\end{abstract} 

\section{Introduction}\label{introduction}

\subsection{Motivation}
The motivation for designing a highly efficient heat dissipation system functioning at ranges above $700\degree$K is for use within energy storage and generation technology. Further in this report, a design will be proposed for the recapture of waste heat from coal-fired power stations to be stored and later used for useful electrical energy generation with the use of high-temperature thermoelectric generator (TEG) devices. The efficiency of such a device is given through the following set of equations \cite{FOM}:

\begin{equation}
 {\eta = \frac{T_H - T_L}{T_H }\frac{\sqrt{1+ZT_{avg}} -1}{\sqrt{1+ZT_{avg}}+ T_L / T_H}}
\end{equation}
\begin{equation}
 {ZT_{device} = \frac{L}{L + 2\rho_c\sigma}ZT_{thermoelement}}
\end{equation}
\begin{equation}
 {ZT_{thermoelement} = \frac{\alpha^2\sigma T}{k}}
\end{equation}

\begin{figure}[h]
\centering
\includegraphics[scale=0.8]{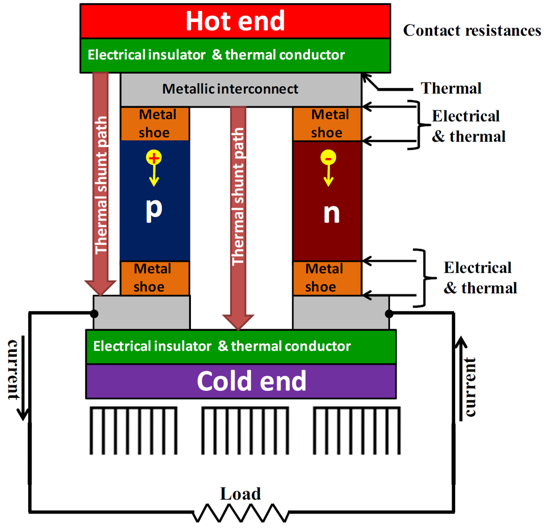}
\caption{Detailed 2D view of TEG with heat dissipation system at bottom modified from \cite{FOM}}
\end{figure}

Notation can be found in the nomenclature. It should also be noted that $k = k_e + k_{Bi} + k_L$. It is clear from the set of above equations that the generation of useful electrical energy through a TEG device can be optimized in many ways such as improving the Seebeck coefficient of each thermoelement in a TEG device or lowering the contact resistance of the components of a thermoelement. Further improvements to all of these parameters can and should be made for IOT to allow TEG devices to be used more widely throughout the industry. However, as can be observed in these reports \cite{coal} \cite{exergy}, a less than 10\% improvement in the heat rate used within wasteful systems such as coal-fired power stations would mean millions in savings annually. The figure of merit (FOM) for a thermoelement as observed in equation three is balanced between the electrical and thermal conductivity of the material used for the thermoelement. Most to all materials used for TEG devices are good conductors both electrically and thermally. This is shown graphically in Figure 1 as the thermal shunt paths. Some improvements have been made to decrease thermal conductivity whilst maintaining electrical conductivity seen here \cite{FOM}, and a FOM of 1.5 correlating with 15\% efficiency for a TEG device was reported in the same paper. Equation 1 shows that the overall efficiency of a TEG device can be improved by raising the hot end temperature and lowering the cold end temperature. Working at high-range temperatures and with materials that must be inherently good thermal conductors means that the cold end temperature of any TEG device will also be quite hot. This report aims to create a heat dissipation system so that the cold end temperature of a high-temperature range TEG device can be efficiently cooled and allow for high efficiencies of the overall TEG device. 

\begin{figure}[h]
\centering
\includegraphics[scale=0.21]{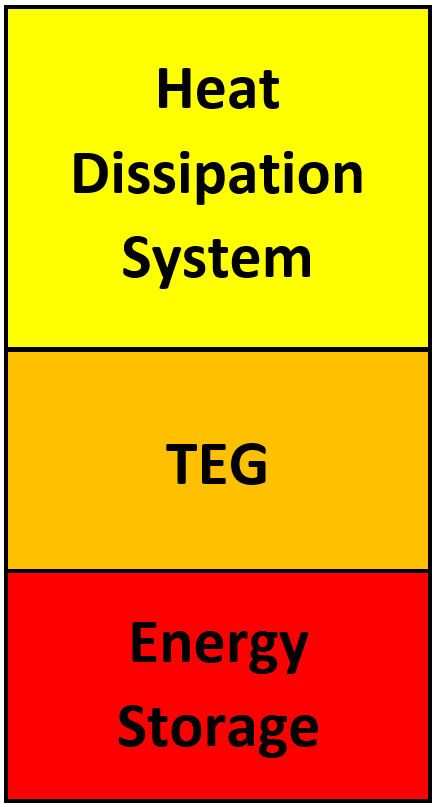}  
\caption{simple 2D view of general setup for power generation system showing heat dissipation system}
\end{figure}

\subsection{Design Problem}
The problem that this study sets to solve is the design of an efficient heat dissipation system for high-temperature applications ($700\degree$ - $1000 \degree$K). Figure 2 shows a general 2D setup for the overall problem to be solved. The study focuses on the yellow component of this diagram. For high-temperature range applications a simple heat sink topology is not efficient enough to effectively dump the heat from the TEG module as the surface area provided using any topology is not large enough \cite{Heat_sinks_TEG} .To overcome this issue a vapour chamber was considered which could effectively transport the heat from the TEG module shown in orange in Figure 2.

\begin{figure}[h]
\includegraphics[scale=0.4]{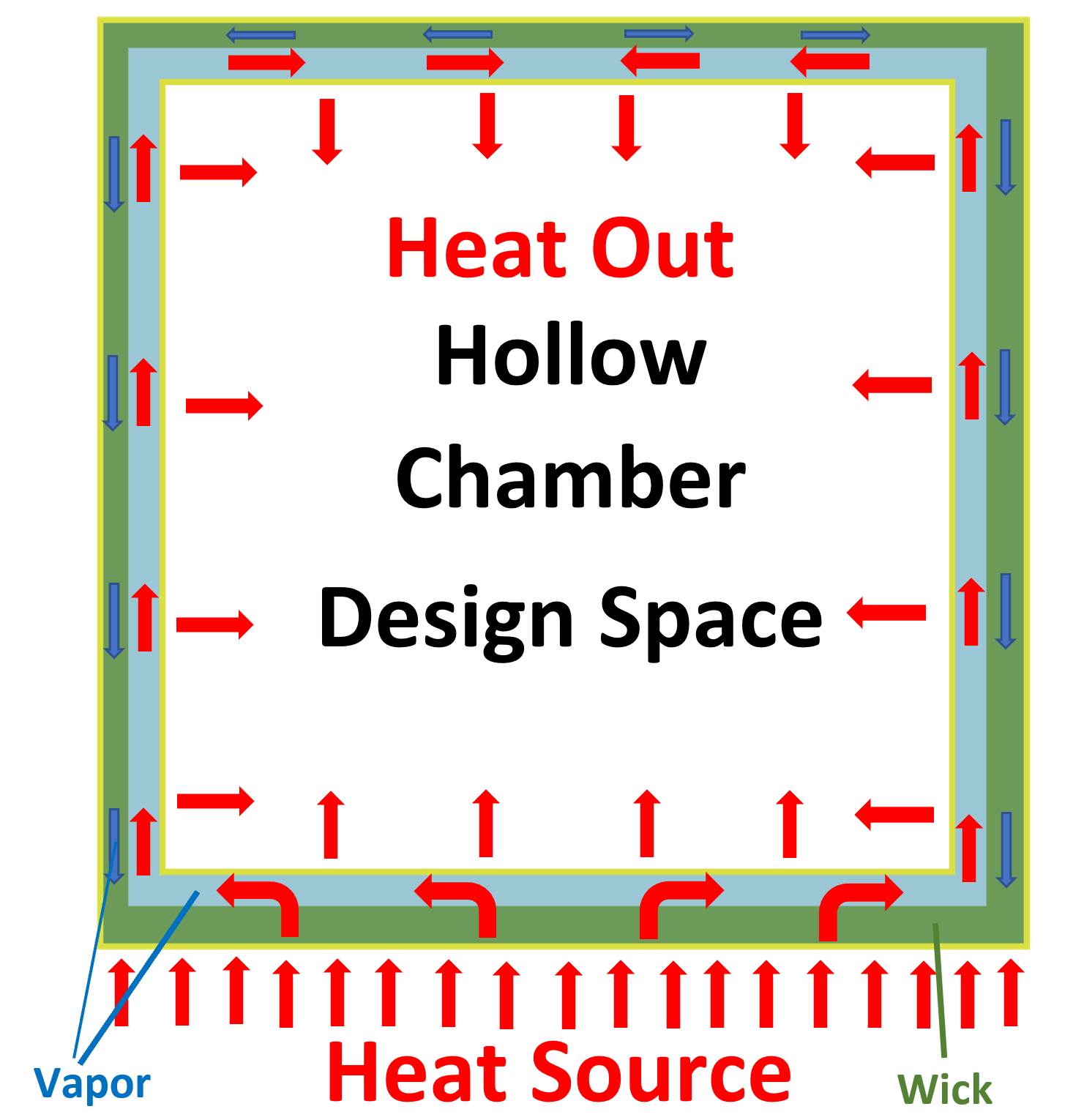}
\caption{Detailed 2D view of vapour chamber showing hollow design space}
\end{figure}

As reported here \cite{Vapour_chamber1}, \cite{Vapour_chamber2} commercially available high-temperature vapour chamber devices are suitable for this application. The intricate design of this vapour chamber i.e. the wick design, vapour and vapour space has not been developed yet for this specific problem but the literature on the current technology shows that it is definitely possible. Assumptions made for using this vapour chamber device were that the steady-state response of the system would effectively transport the heat from the top of the TEG module creating a homogeneous vapour temperature across all walls of the inside of the hollow vapour chamber device. The proposed hollow chamber device is shown in Figure 3. In considering the possible 2D and 3D geometrical parameters of the 3D figure used for the design of the heat dissipation system, a contributing factor to the design of the geometry was the surface area to volume ratio. A high volume was needed to increase the surface area available within the inner mesh to be created within the design space. A high surface area was also needed to allow efficient heat transfer from the vapour chamber to the inner mesh to allow for the forced convection cooling to take heat away efficiently. Figure 4 shows that the obvious choice to achieve the highest surface area to volume ratio is a tetrahedron structure. However, this geometry introduced far more complexity to the design of forced convection cooling promising to lower the turbulent inflow boundary condition. The tetrahedron design also did not interface well with the TEG module. For these reasons, a cube structure was decided upon for the housing of the vapour chamber and inner mesh. Figure 5 shows the final design chosen for the hollow cuboid vapour chamber.

\begin{figure}[h]
\centering
\includegraphics[scale=0.15]{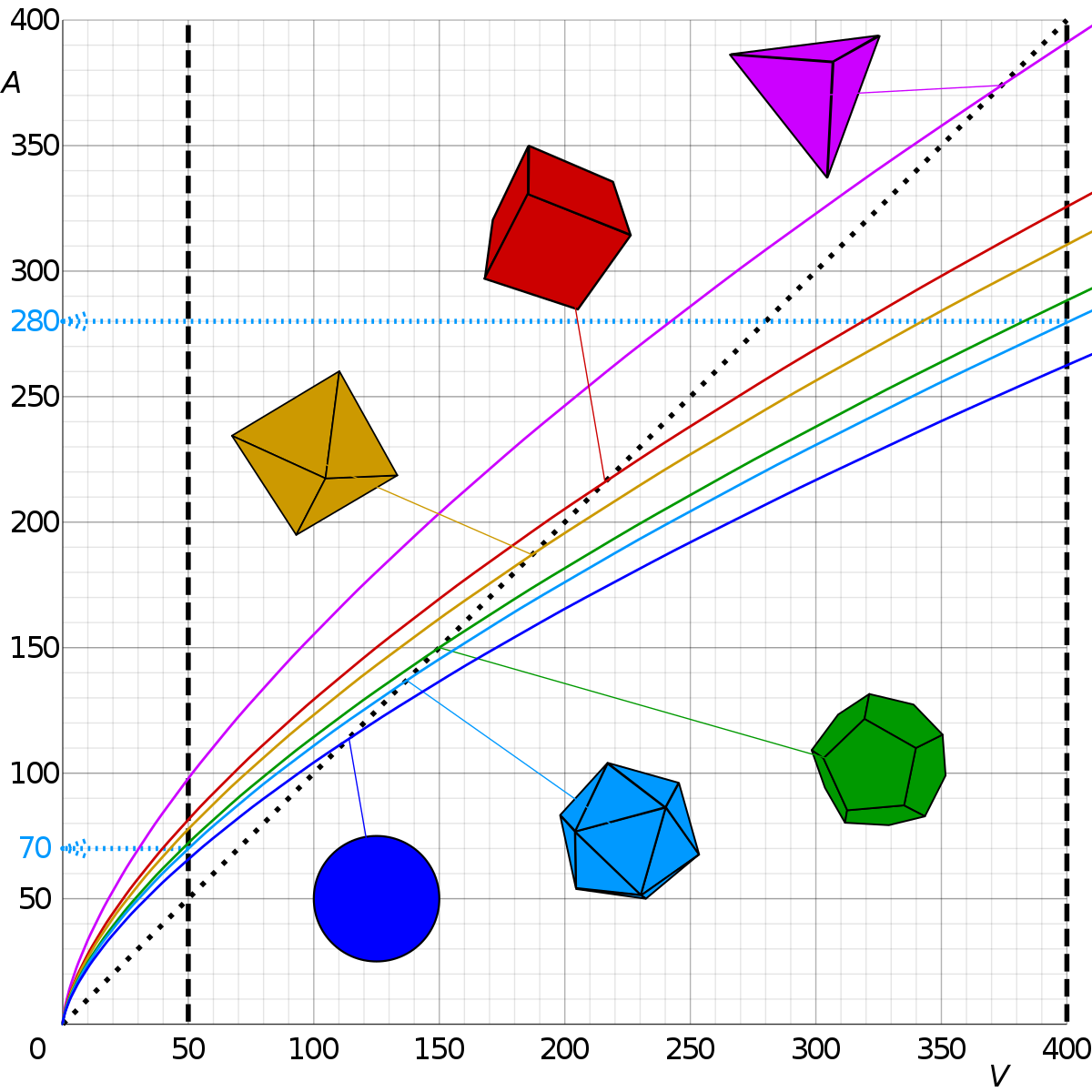}
\caption{Comparison of surface area vs volume of shapes}
\end{figure}

\begin{figure}[h]
\includegraphics[scale=0.7]{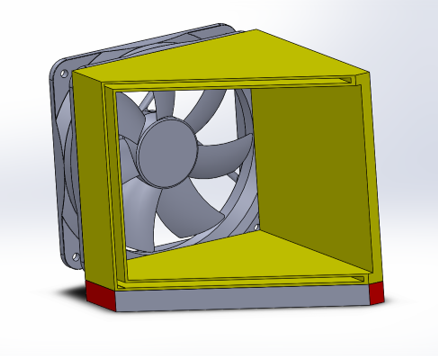}
\caption{Detailed 3D view of the hollow cuboid vapour chamber. Note that a cut has been taken out of the chamber to show the inner vapour space. The TEG module is shown below in red. }
\end{figure}

\section{Methodology}\label{methodology}
\subsection{Governing Equations}
The optimisation problem for both the parametric and topology-based approaches follows the following form

\begin{equation}
  \begin{aligned}
    \text{minimise} \quad f_{obj}(\Omega)  \\
    \text{subject to} \quad f_{con} \leq0  \\  
    \Omega_j^{min} \leq \Omega_j \leq \Omega_j^{max} , \quad j = 1,\dots , n
  \end{aligned}
\end{equation}

where $f_{obj}(\Omega)$ is the objective function which for this case will be the temperature gradient across the design space, $f_{con}$ are the design constraints for the particular optimisation problem, and $\Omega$ is a vector is $n$ design variables $\Omega_j$ which for this case will be the mesh points in the design space. \cite{Lange_paper}
\newline
The turbulent flow conjugate heat transfer module was utilised in COMSOL which combines both the heat transfer in solids and fluids and the turbulent flow fluid flow modules. The steady-state pure heat conduction Fourier law governs the heat transfer for the numerical solutions found:

\begin{equation}
 {-\div(k\grad T) = Q}
\end{equation}

A fluid incompressibility constraint was used for the water within the design domain shown in Equation 6. The Naiver-Stokes fluid flow equation was used to govern the movement of the water in the design space. 

\begin{equation}
 {\div(\rho U) = 0}
\end{equation}
\begin{equation}
 {\rho (U \cdot \grad U) = - \grad P + \div[\eta_f(\grad U + (\grad U )^T)] - \alpha_{por} U}
\end{equation}

The turbulent flow and heat transfer physics were then combined with COMSOL multi-physics using a convection-diffusion heat transfer equation
\begin{equation}
 {-\rho C (U \grad T ) = \div(k \grad T) + Q}
\end{equation}

\subsection{Objective Function}
The objective function aims to minimise the temperature gradient across the design space as fast as possible so that the heat is evenly distributed across the maximum amount of surface area to be carried away by forced convection. The objective function was therefore defined as the integration of the divergence of the temperature of the design space. The gradient of the temperature was squared to minimise the error in the gradient operator. 

\begin{equation}
 {f_{obj}(\Omega) = \int_{\Omega}^{} k(\grad T)^2 d\Omega}
\end{equation}
\subsection{Parametric Optimisation}
Eight similar setups were para-metrically swept through for the minimisation of the objective function. 
\begin{figure}[h]
\centering
\includegraphics[scale=0.3]{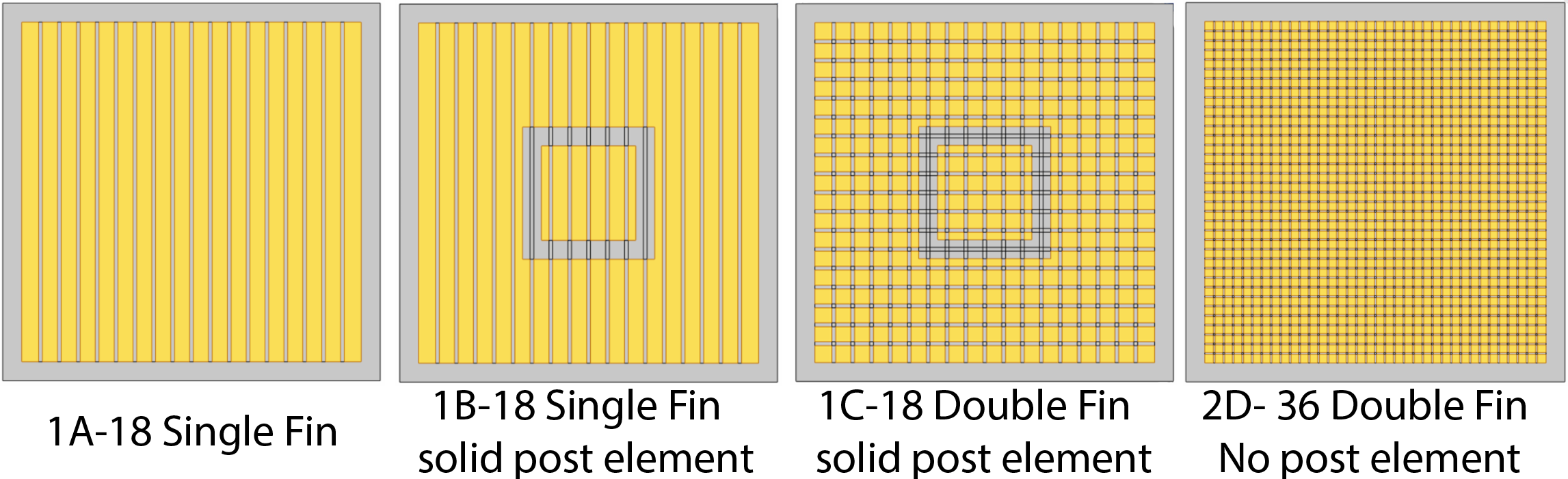}
\caption{Showing 4 of the 8 setups para-metrically swept for optimisation. Setup 1 utilised 18 fins. Setup 2 utilised 36 fins. A solid post element was integrated to the middle for setup B and C}
\end{figure}
\begin{figure}[h]
\centering
\includegraphics[scale=0.25]{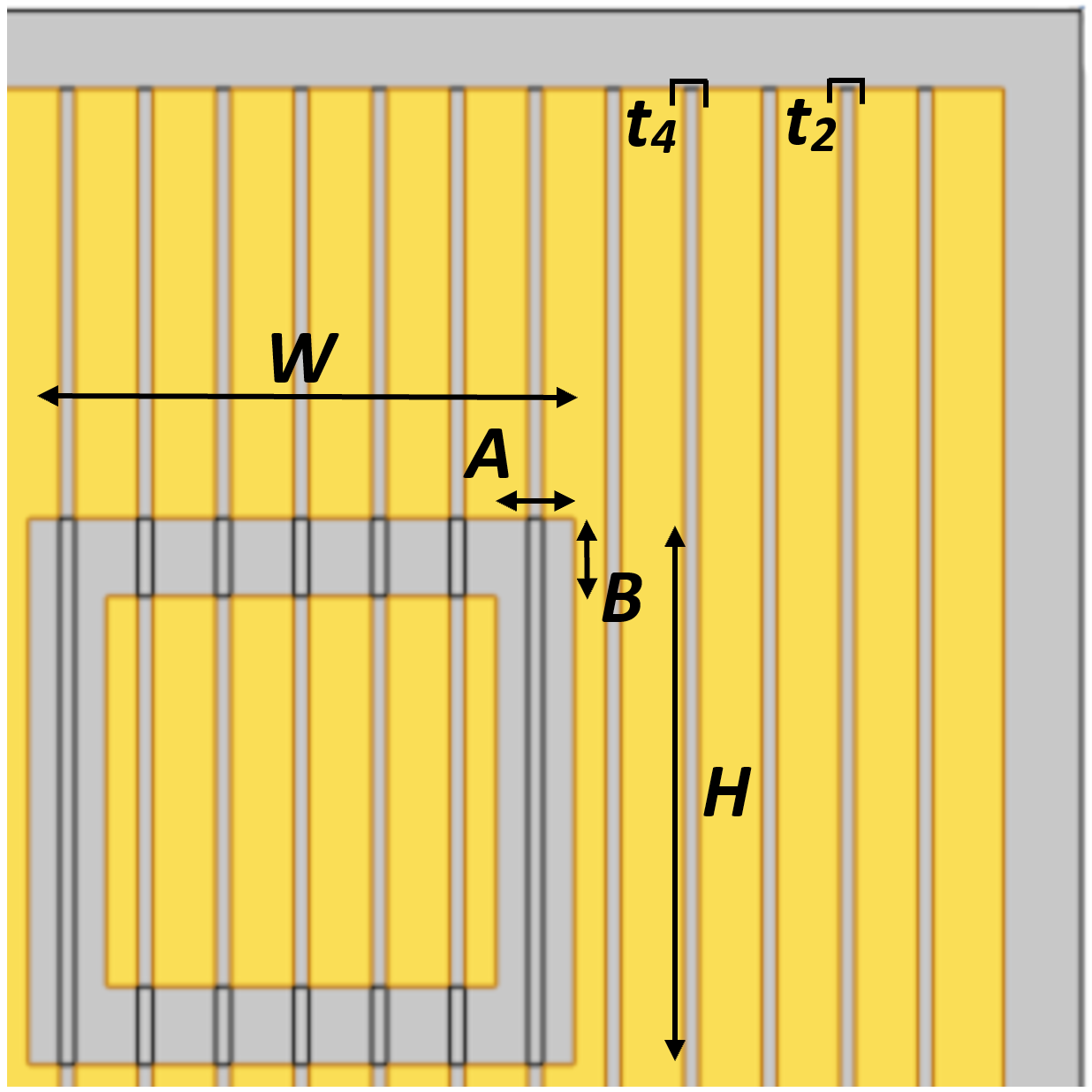}
\caption{Diagram showing fin thickness, $t$, post height, width and thickness, $H$, $W$, $A$, $B$ for parametric sweep. }
\end{figure}

It was determined that fin structures between 18 and 36 performed the best through a series of tests. Smaller numbers of fins did not allow for the temperature to reach the centre fast enough and higher several fins impeded the turbulent flow too greatly for the forced convection to work efficiently. To reduce computational processing time the two ends of the spectrum were para-metrically swept. It should be noted however that some fins between 18 and 36 would also work effectively. This is detailed further in the results and analysis section. The solid posts were added for testing in an attempt to draw heat from the sides in towards the centre as initial testing showed that the heat tended to remain on the outside of the heat dissipation mesh. Five parameters were para-metrically swept through in pursuit of an optimal setup. The first was fin thickness, $t$. An individual fin thickness was assigned to each thin meaning that 18 fin thickness parameters were used for setup 1A and 72 fin thickness were used for setup 2D. The second and third were the height and width, $H$ and $W$, of the solid post element in the middle of setup B and C. The last parameters were the thickness of the solid post element for setup B and C in the X and Y direction, $A$ and $B$. Figure 7 illustrates all parameters that were para-metrically swept through.

\subsection{Topology Optimisation}
The material distribution method was used to determine the optimal topology for the given design space. A control variable, 
$\theta_c$ was introduced which was bounded between 0 and 1. $\theta_c$ = 1 corresponds to a solid material whereas $\theta_c$ = 0 corresponds to no material. A density model feature was utilised in determining the optimal topology. The density model introduced a minimum length scale using a Helmholtz filter radius, $R_{min}$. 

\begin{equation}
 {\theta_f = R_{min}^2 \grad^2 \theta_f + \theta_c}
\end{equation}

\begin{figure}[h]
\centering
\includegraphics[scale=0.35]{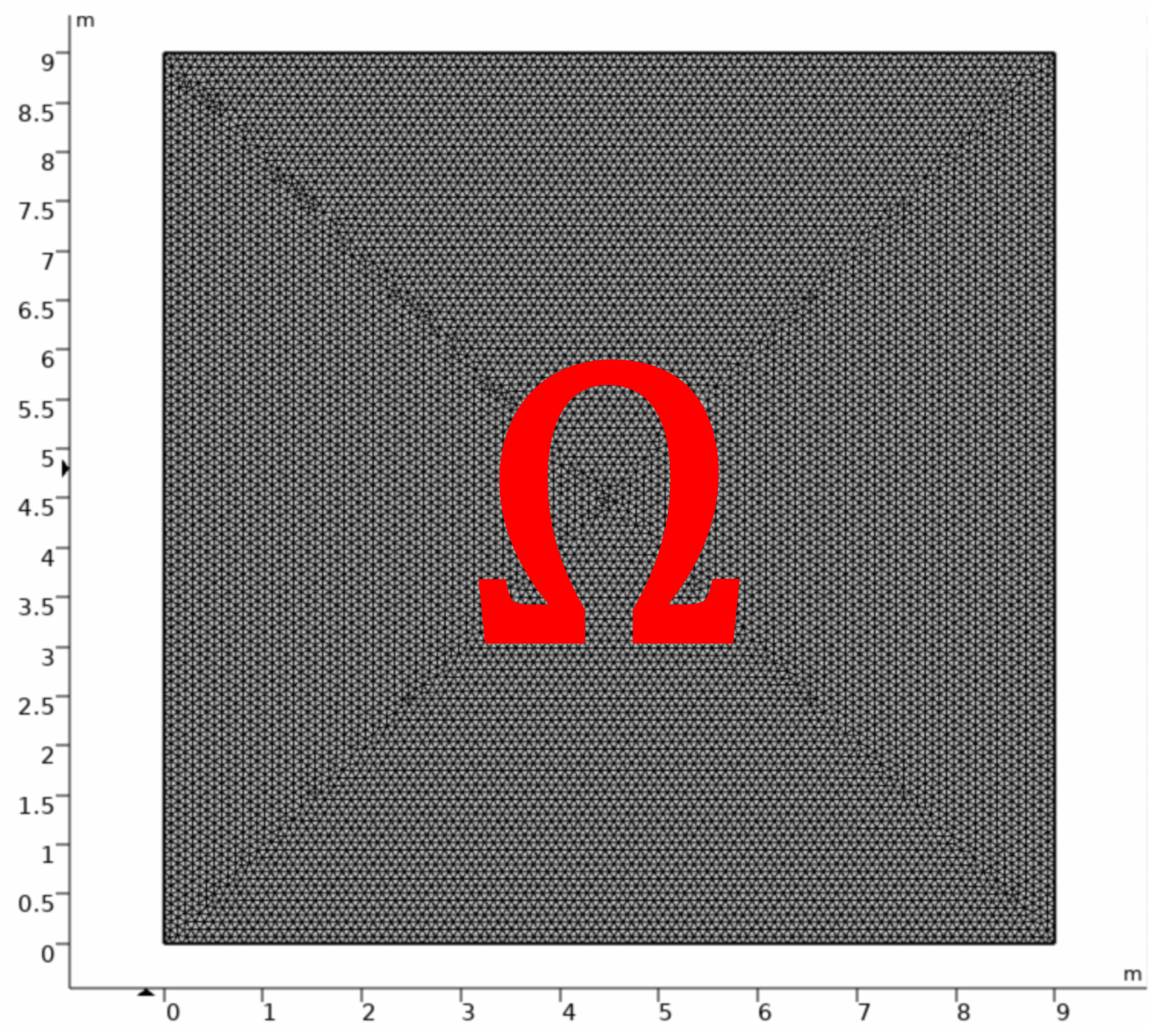}
\caption{Diagram showing the 2D design space mesh. Triangular tetrahedrons were used as the tessellation pattern with extremely fine mesh size. Each triangle represents a node that corresponds to some density value between $\theta_f$ and 1 which further correlates to a specific SIMP value for thermal conductivity and heat capacity.}
\end{figure}
A simple Isotropic Material with a Penalisation method was introduced for the material density, heat capacity and thermal conductivity over the design space as well as in the objective function. 
\begin{equation}
 {k_{SIMP} = k_1 + \theta_p(k_2 - k_1)}
\end{equation}
\begin{equation}
 {C_{SIMP} = C_1 + \theta_p(C_2 - C_1)}
\end{equation}
where $\theta_p$ is the interpolation SIMP exponent. The interpolation SIMP exponent works to assign a relative density to each element in a design mesh between $\theta_f$ and 1. $k_1$, $k_2$, $C_1$ and $C_2$ are the thermal conductivity and specific heat capacity for solid material used and water \cite{Topology_Webpage}. The objective function is altered once again to allow for an active forced convection cooling with water to take place over the topology. The derivation for this objective function can be found here \cite{Lange_paper}

\begin{equation}
  \begin{aligned}
    f_{obj,top}(\Omega) = (1-q)\int_{\Omega}^{} k_{SIMP}(\grad T)^2 d\Omega \\ + q\frac{h_0h_{max}}{A}\int{\Omega}{\abs{\rho_{des}(\Omega)}^2d\Omega}
  \end{aligned}
\end{equation}

\subsection{Implementation}
The assumption that was made with regard to the hollow cuboid vapour chamber was that at its steady-state response, it would homogeneously disperse the heat flux from the bottom receiving surface to all surfaces on the inside of the hollow cube. With this in mind, the temperature boundary conditions for the simulation were set so that a temperature of $1000 \degree$K was present on all of the inside surfaces of the vapour chamber. An out-of-plane radiation was set to the outside boundaries of the vapour chamber to simulate the convection cooling and heat transfer from the metallic heat dissipation surface to the surrounding ambient air temperature. An out-of-plane heat flux was also included to simulate the forced convection cooling that will be present in the 3D simulation seen later in this report. Structural steel was used as the material for the vapour chamber and mesh as it has a melting point well above $1000 \degree$K. A water material was then added to the inside of the vapour chamber at ambient temperature and a flow condition was added to simulate the extraction of heat from the system through a forced convection cooling system utilising water as the transfer medium. This out-of-plane heat flux was only included on the water domain so heat taken out of the system by forced convection was only done so through the water domain. Figure 9 shows the initial response of the hollow design space filled only with water and applying these boundary conditions.

\begin{figure}[h]
\centering
\includegraphics[scale=0.75]{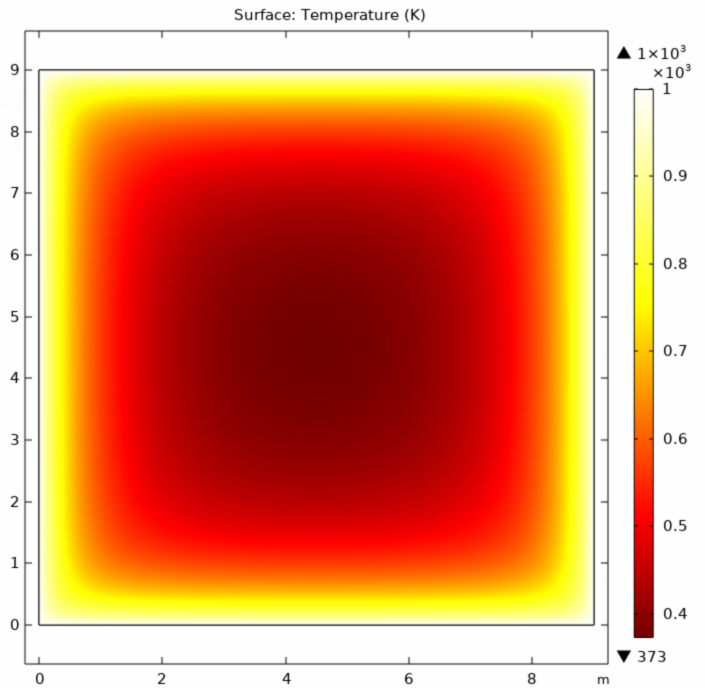}
\caption{Response of the hollow design space filled only with water to the boundary conditions as described in the Implementation section}
\end{figure}

\section{Results}\label{results}
\subsection{Number of Fins}
A MATLAB-COMSOL script was written to conduct testing to determine the optimal number of fins within the design space for the parametric testing to be conducted. Both the pair-fin design wherein fins are placed both vertically and horizontally seen in designs 1-2D (refer to Figure 6) as well as the single-fin design wherein fins are only placed horizontally (designs 1-2 A and B, refer to Figure 6) were used to determine the optimal number of fin numbers. The number of fins and pairs of fins was increased from 0 - 80 by 2 at each iteration. The fin size was also decreased at each iteration to account for the shrinking design space size. Figure 10 shows that both the single and paired fin topology have a minimum of 36 fins. The variance of the objective function between 18 and 36 fins is quite small. Figures 11 and 12 both show that fringing effects begin to occur at 36 fins and are quite prominent as the number of fins increases from there. These fringing effects wherein heat builds up on the outer layer of the mesh are less prominent at 18 fins. For these reasons, both the 18 and 36 fin topology as seen in figures 11 and 12 were tested para-metrically. 
\begin{figure}[h]
\centering
\includegraphics[scale=0.45]{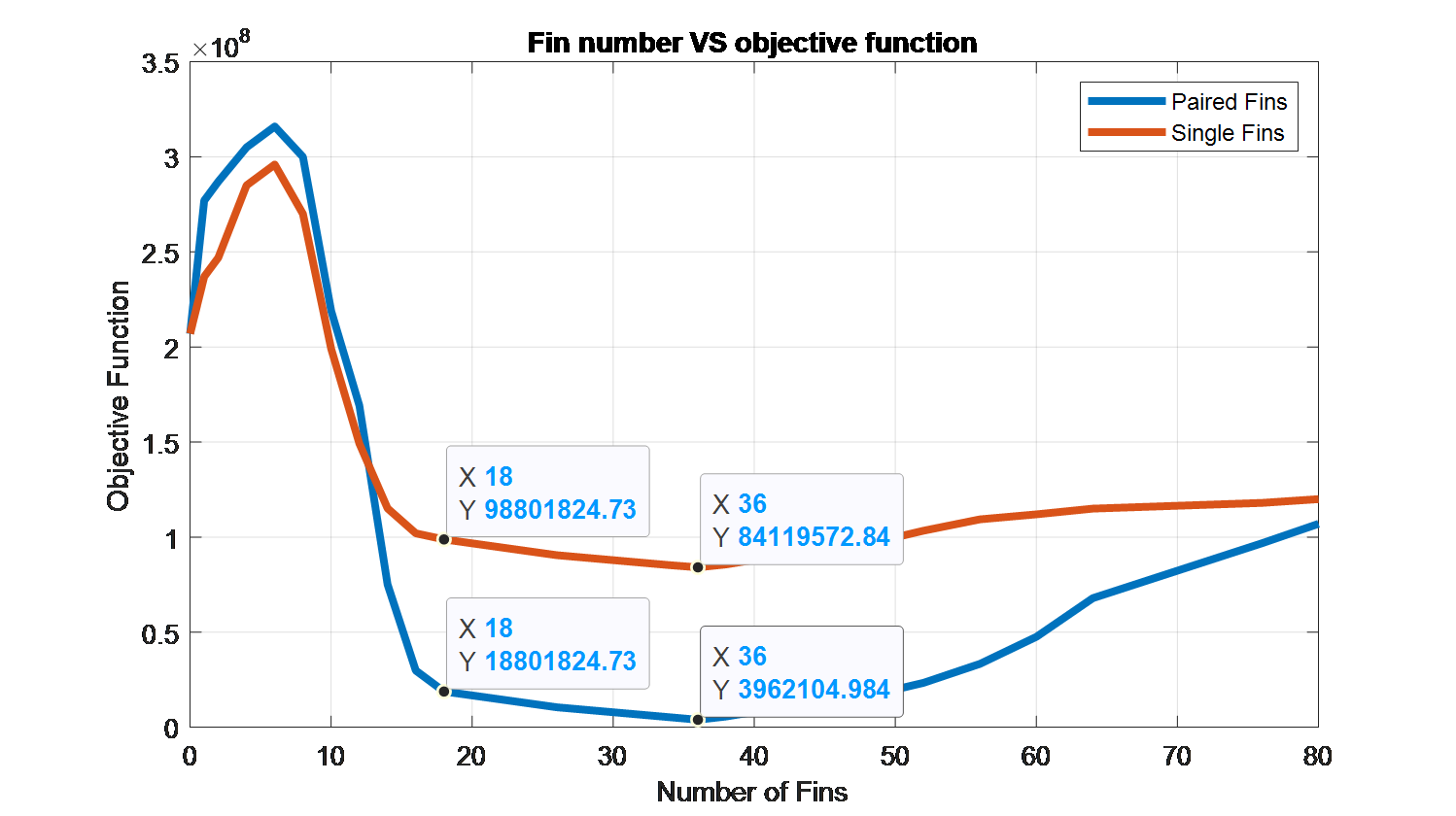}
\caption{Plot for a number of fins VS objective function for single and paired fin topology.}
\end{figure}

\begin{figure}[h]
\centering
\includegraphics[scale=0.2]{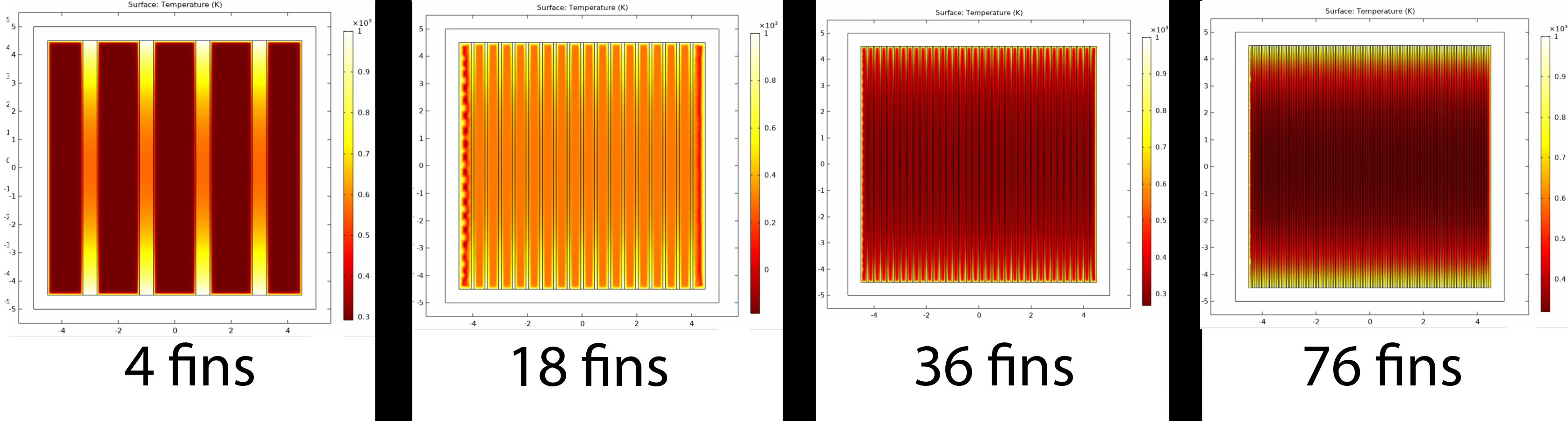}
\caption{Plot of Heat distribution for single fin topology for 4,18,36 and 76 fins. See appendix for larger images}
\end{figure}

\begin{figure}[h]
\centering
\includegraphics[scale=0.2]{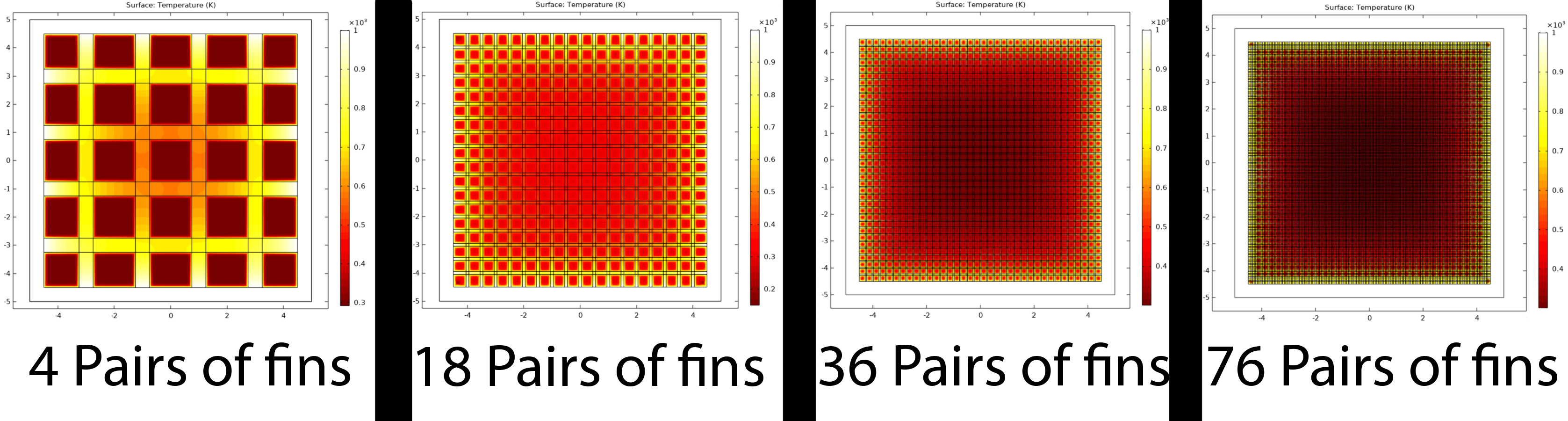}
\caption{Plot of Heat distribution for paired fin topology for 4,18,36 and 76 paired fins. See appendix for larger images}
\end{figure}

\subsection{Parametric Optimisation}
The results of the parametric and topology-based optimisation are shown in Table 1. Initially, it seems that the 36-fin topology yields a lower objective function. Through further testing though it is observed that the larger fin number setup shows little improvement through parametric optimisation. Setup 2D shows the greatest improvement in the objective function for the 36 fin topology with only a variance of $10.8 \times 10^{5}$. This is contrasted vastly with the improvement in the 18-fin setup. An improvement of $828.4 \times 10^{5}$ from the parametric optimisation of setup 1C yielded the lowest objective function of all the parametric optimisation tests conducted. It was also observed that the single-finned setup yielded a larger objective function for both the 18 and 36-fin setup which correlates well with the data found in Figure 10. Overall, the topology optimisation gave the lowest objective function with an optimal result of $9.86\times 10^{5}$ which was only $4.74\times 10^{5}$ lower than the lowest result from the parametric testing. This is explored further in the topology optimisation results section. 

\begin{table}[ht]
\centering
\caption{Test setup VS Initial objective function VS final objective function post-processing.}
\label{table:justsheepcnn}
\begin{tabular}{ccc}
\toprule
Setup & Objective Initial($\times 10^{5}$) & Objective Final($\times 10^{5}$) \\
\midrule
Control & NA    & 2070 \\ 
1A      & 980   & 39.9 \\  
1B      & 980   & 42.5 \\
1C      & 843   & 14.6 \\
1D      & 840   & 14.9 \\
2A      & 180   & 84.9 \\
2B      & 106   & 85.2 \\
2C      & 39.6  & 35.4 \\
2D      & 39.1  & 28.3 \\
Topo1   & NA    & 9.86 \\
Topo2   & NA    & 26.6 \\
\bottomrule
\end{tabular}
\end{table}

Figures 13 and 14 show the temperature distribution before optimisation and figure 15 and 16 show the same post optimisation.

\begin{figure}[h]
\centering
\includegraphics[scale=0.5]{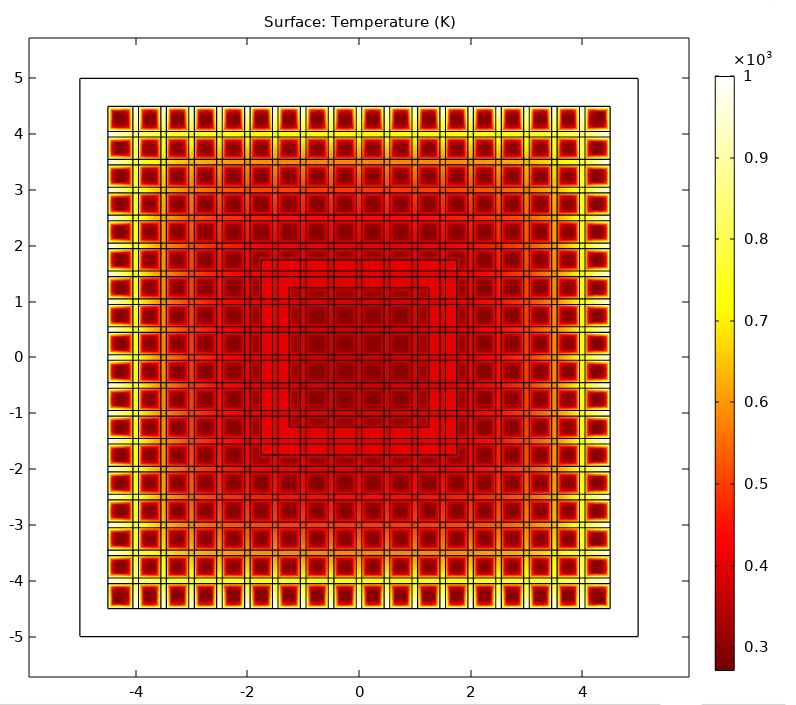}
\caption{Paremetric Results: Plot of Heat distribution for setup 1C before parametric optimisation}
\end{figure}

\begin{figure}[h]
\centering
\includegraphics[scale=0.5]{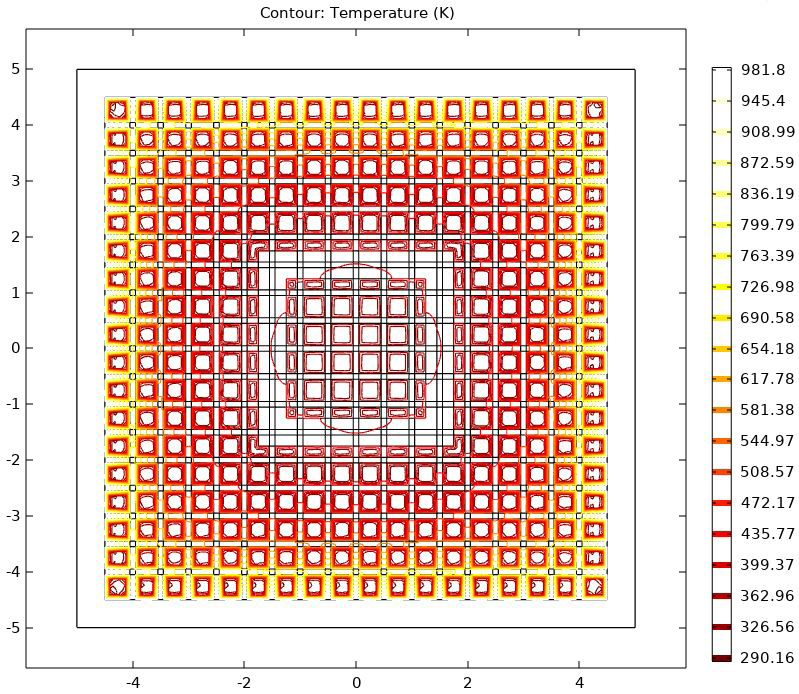}
\caption{Paremetric Results: Plot of Iso-thermal contours for setup 1C prior to parametric optimisation}
\end{figure}

\begin{figure}[h]
\centering
\includegraphics[scale=0.5]{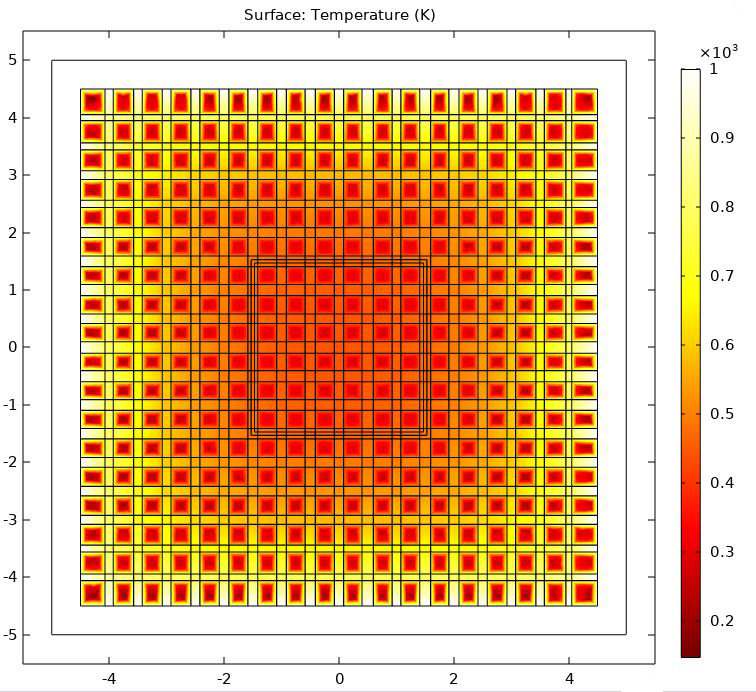}
\caption{Paremetric Results: Plot of Heat distribution for setup 1C post parametric optimisation}
\end{figure}

\begin{figure}[h]
\centering
\includegraphics[scale=0.5]{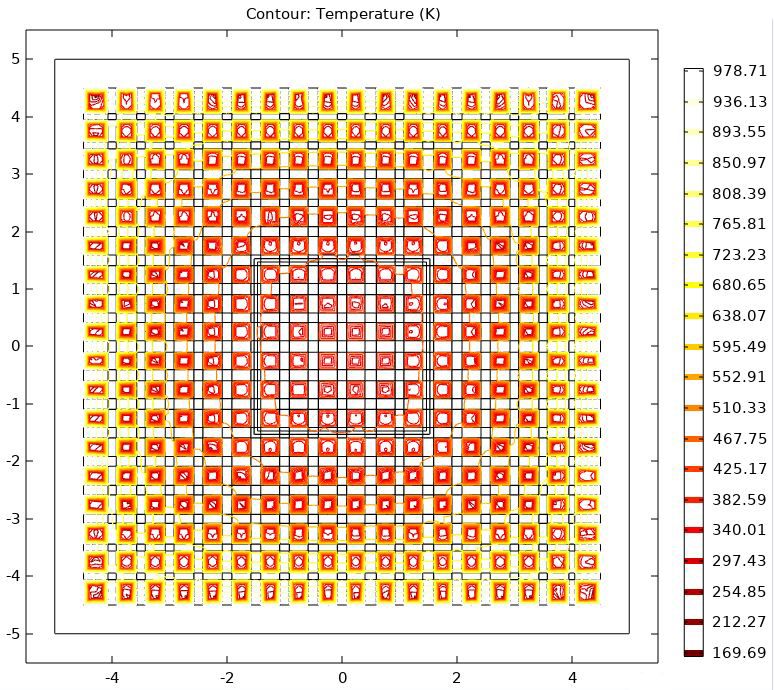}
\caption{Paremetric Results: Plot of Iso-thermal contours for setup 1C post parametric optimisation}
\end{figure}

\begin{figure}[h]
\centering
\includegraphics[scale=0.45]{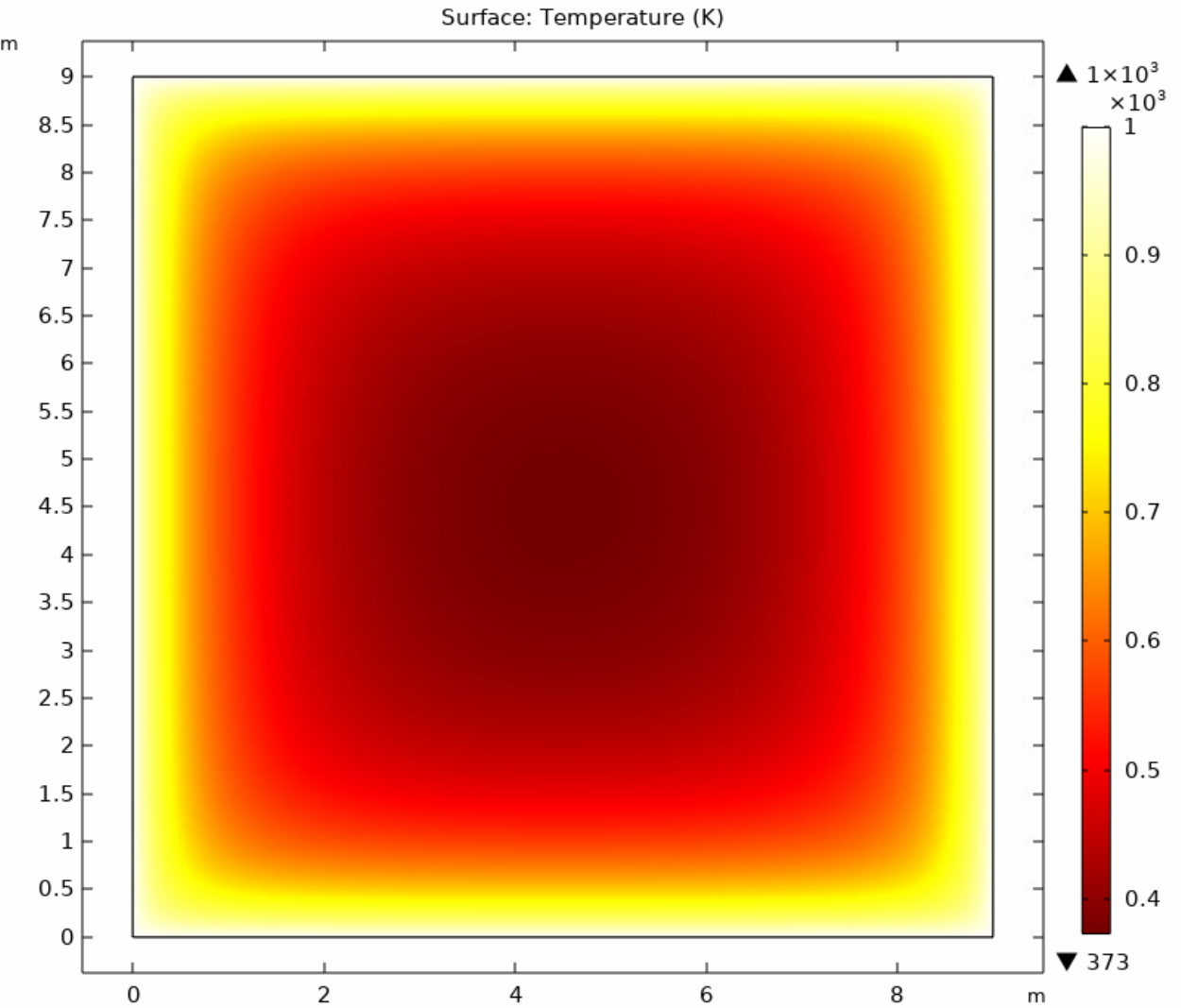}
\caption{Density method results: Heat distribution for the response of the non-optimised system}
\end{figure}

\begin{figure}[h]
\centering
\includegraphics[scale=0.45]{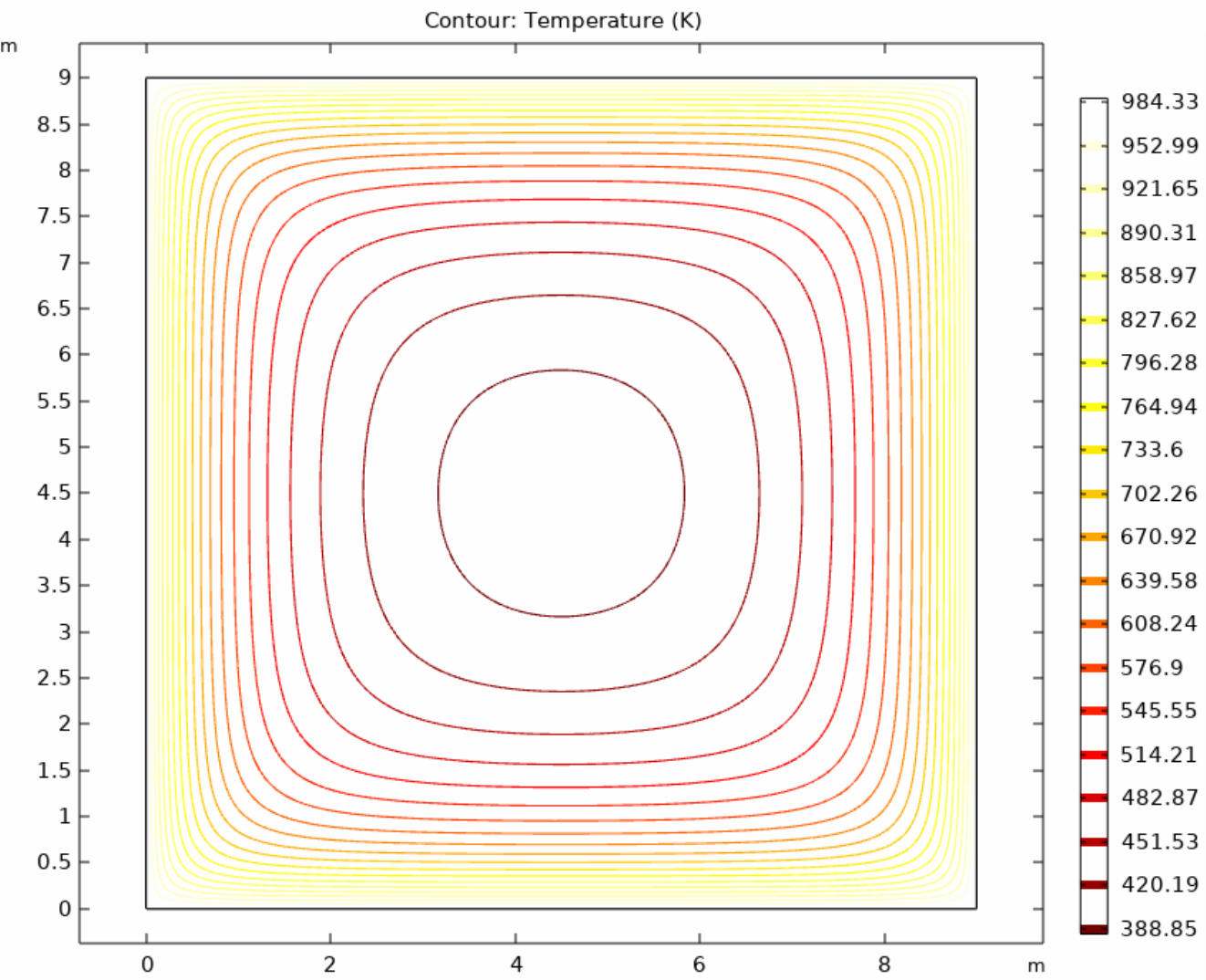}
\caption{Density method results: Iso-thermal contours for the response of the non-optimised system}
\end{figure}

\begin{figure}[h]
\centering
\includegraphics[scale=0.45]{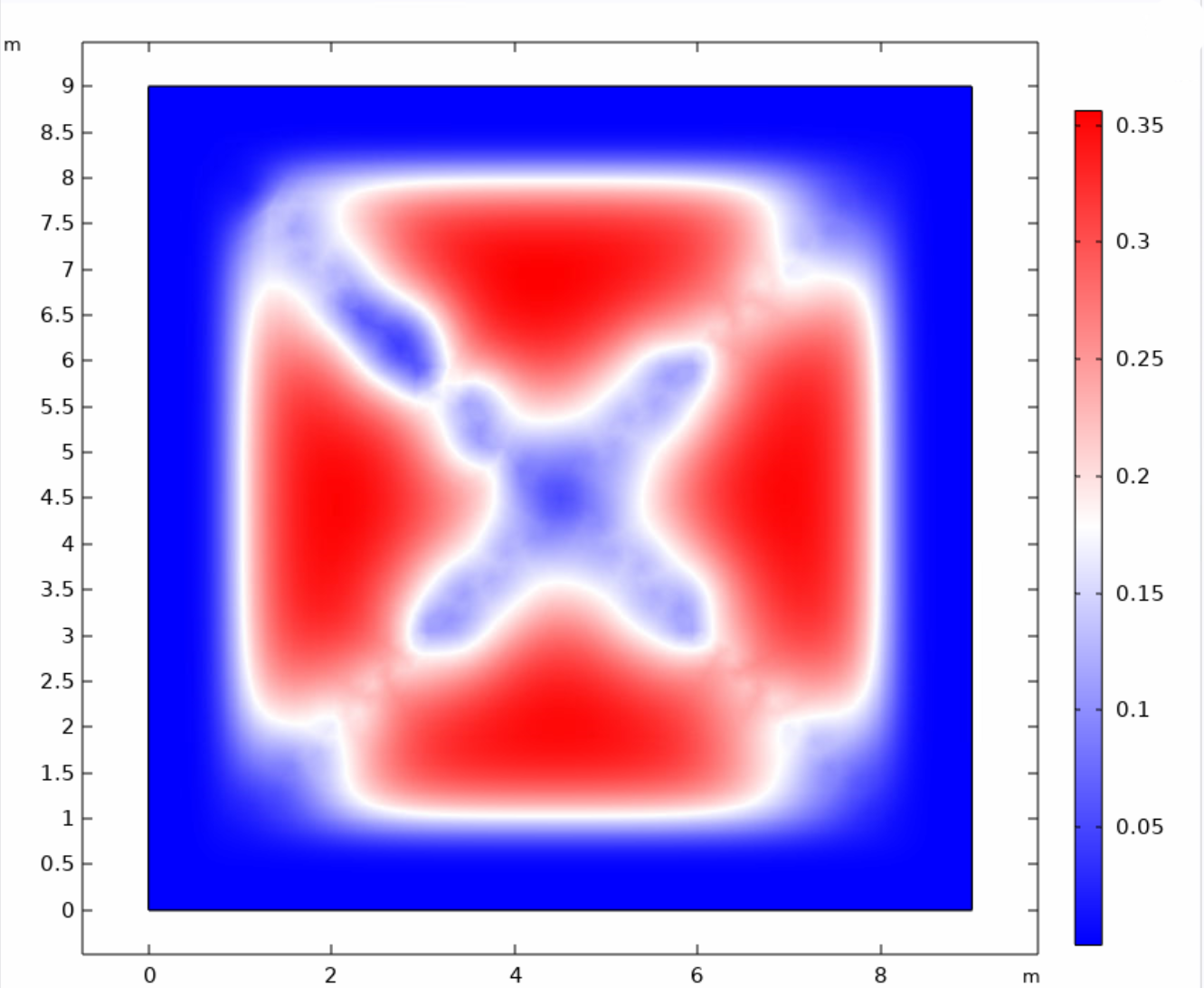}
\caption{Density method results: Result of topology optimisation}
\end{figure}

\begin{figure}[h]
\centering
\includegraphics[scale=0.45]{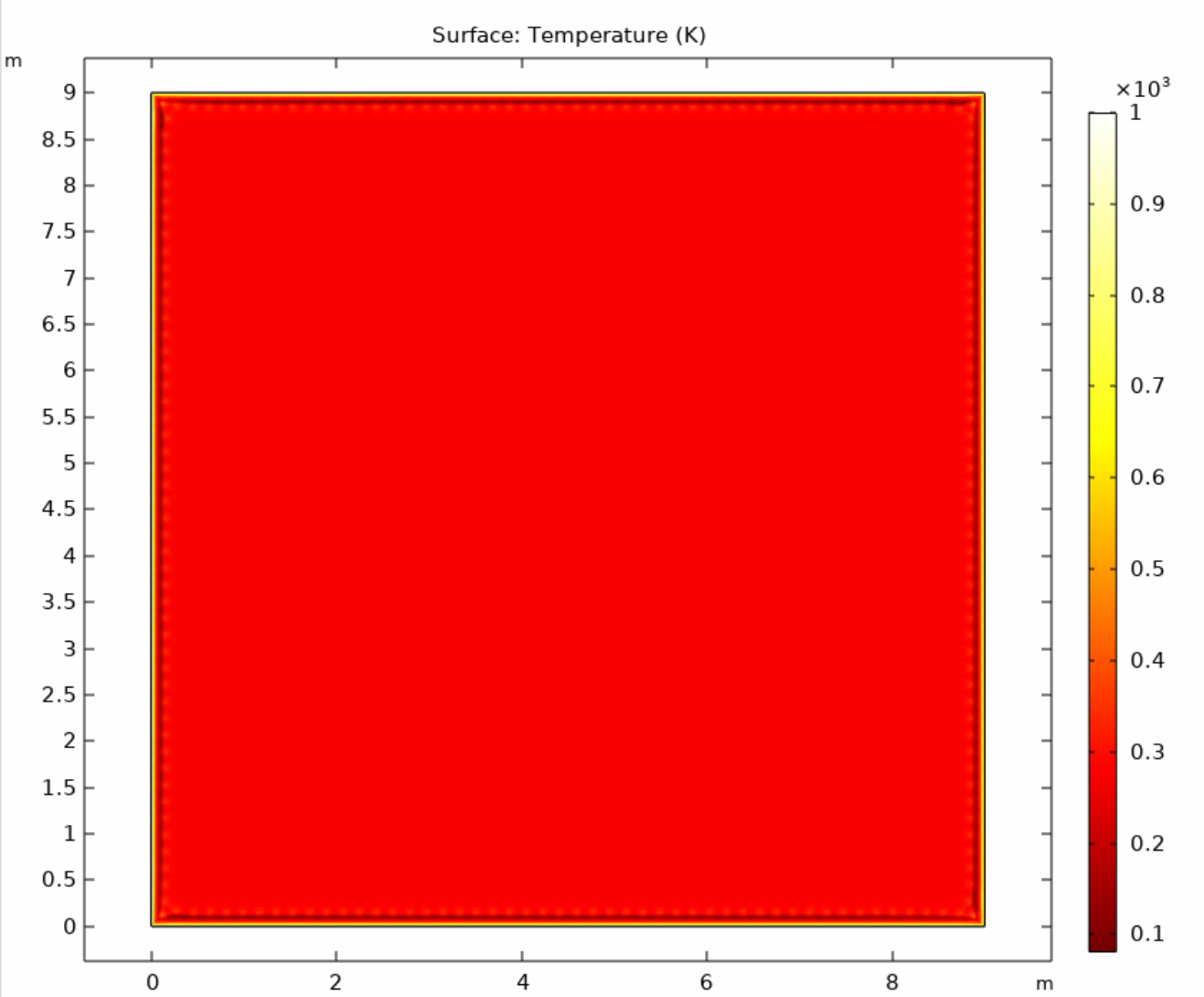}
\caption{Density method results: Heat distribution for the response of the non-optimised system}
\end{figure}

\begin{figure}[h]
\centering
\includegraphics[scale=0.45]{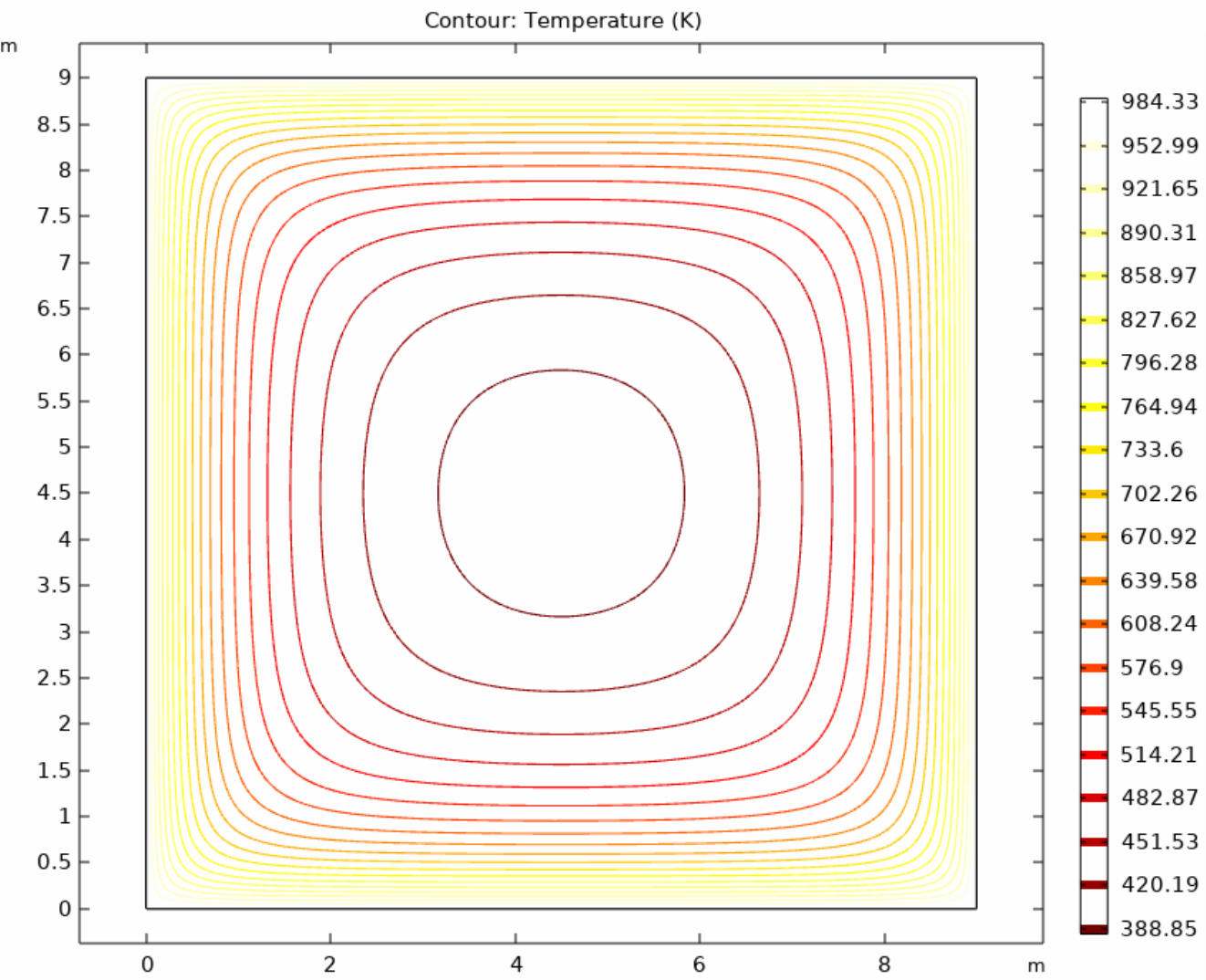}
\caption{Density method results: Iso-thermal for the response of the optimised system}
\end{figure}
\begin{figure}[h]
\centering
\includegraphics[scale=0.45]{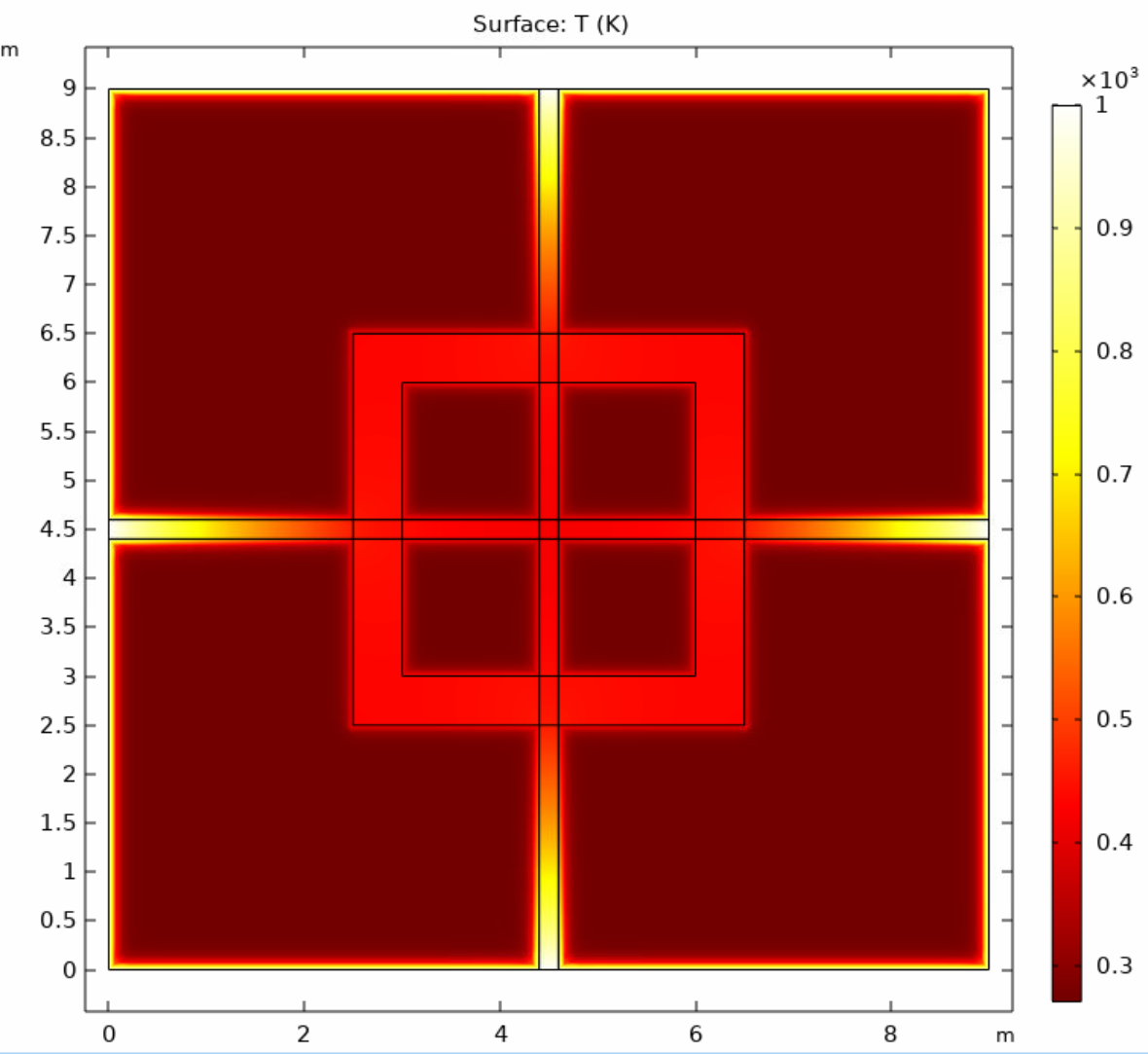}
\caption{Heat distribution for the response of the non-optimised system design 2}
\end{figure}

\begin{figure}[h]
\centering
\includegraphics[scale=0.45]{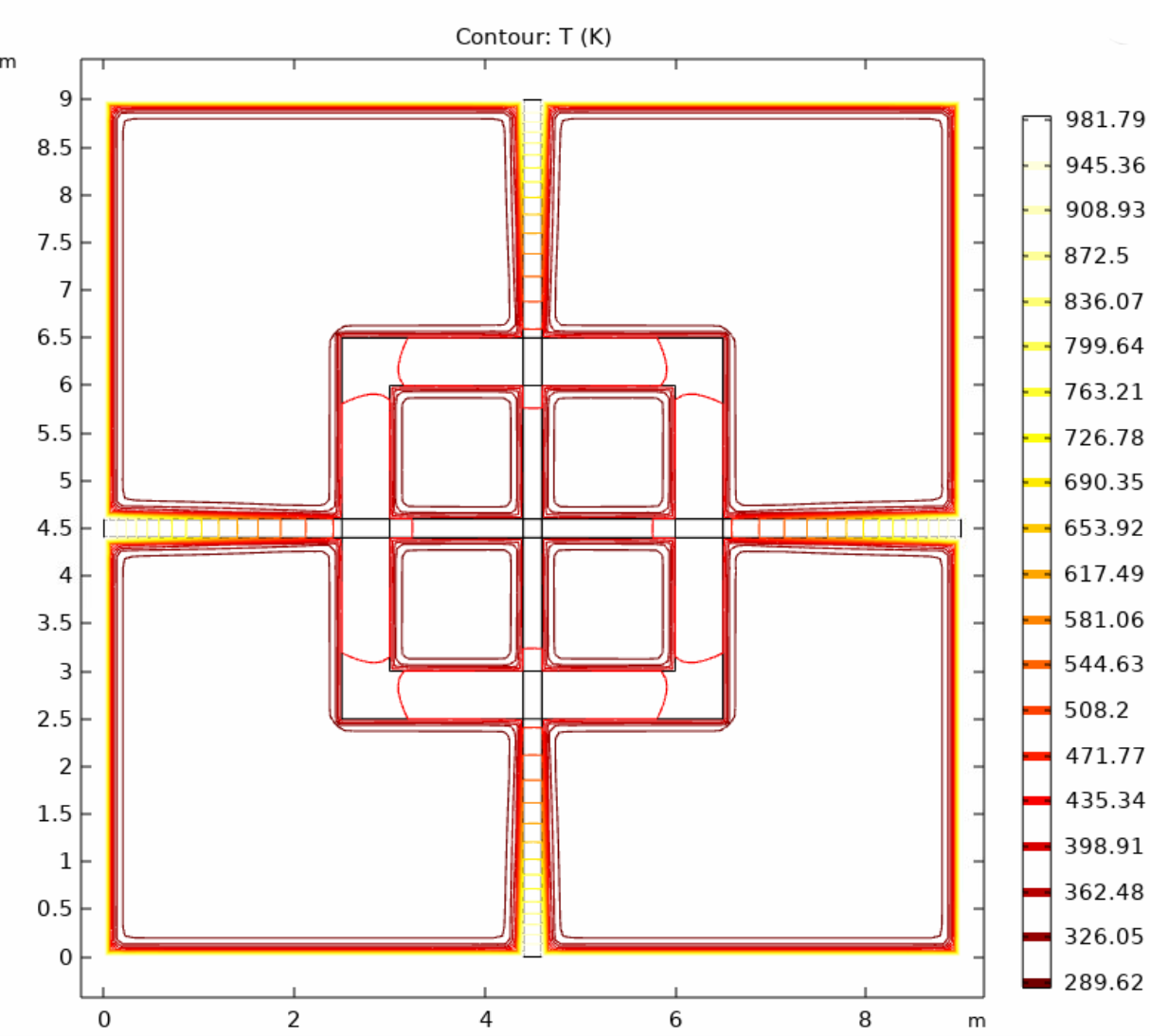}
\caption{Iso-thermal contours for the response of the non-optimised system design 2}
\end{figure}

\begin{figure}[h]
\centering
\includegraphics[scale=0.45]{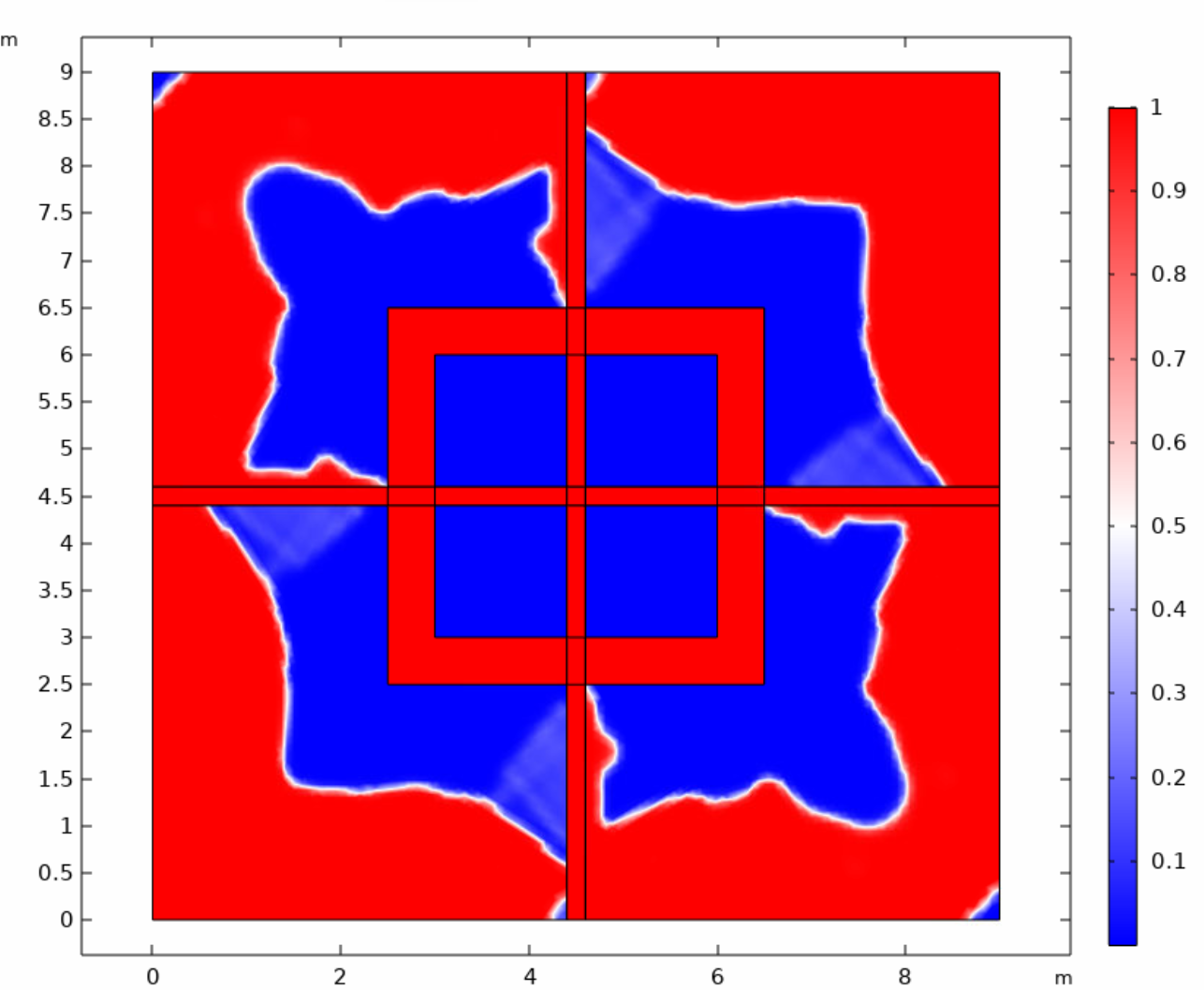}
\caption{Result of topology optimisation design 2}
\end{figure}

\begin{figure}[h]
\centering
\includegraphics[scale=0.45]{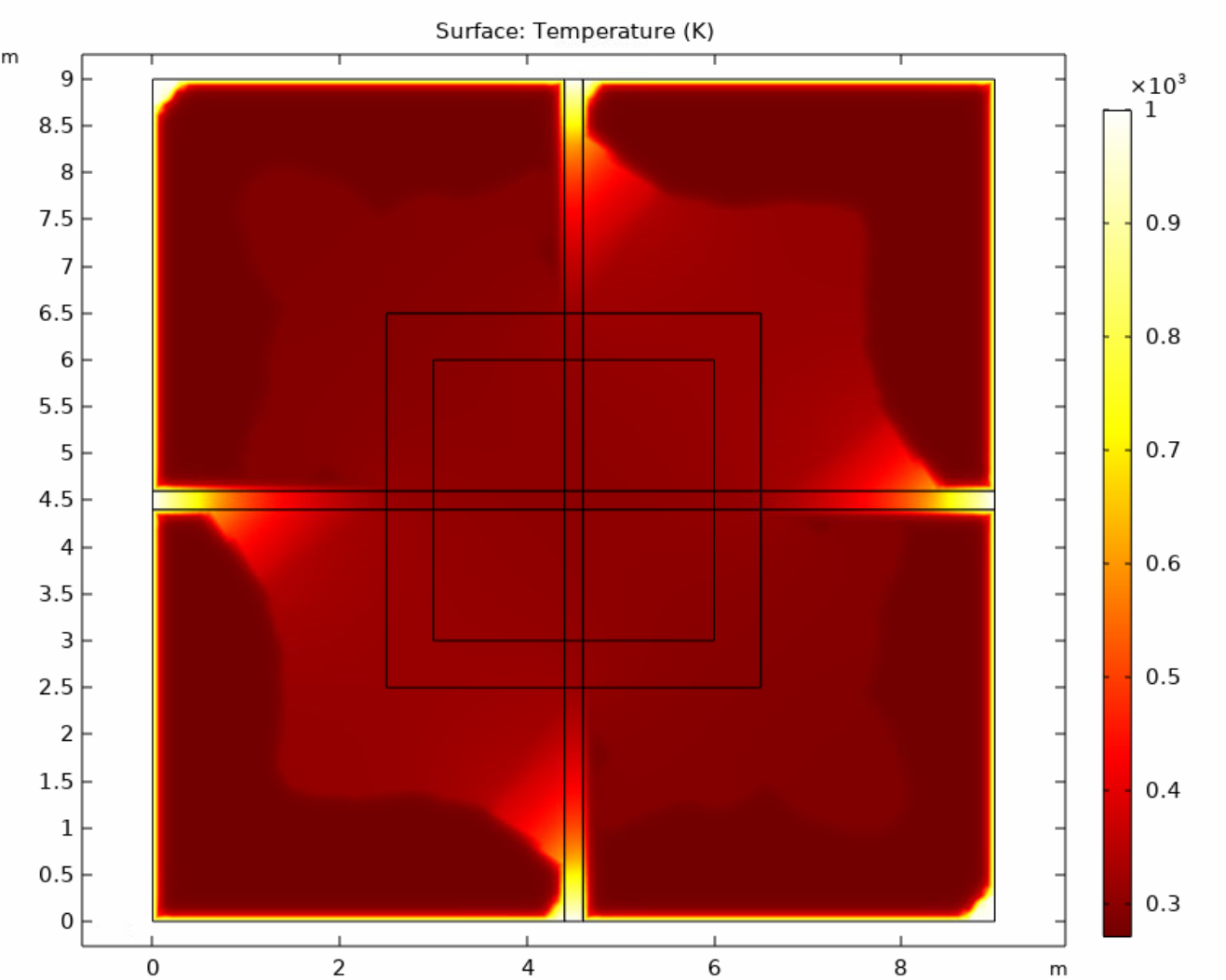}
\caption{Heat distribution for the response of the non-optimised system design 2}
\end{figure}

\begin{figure}[h]
\centering
\includegraphics[scale=0.45]{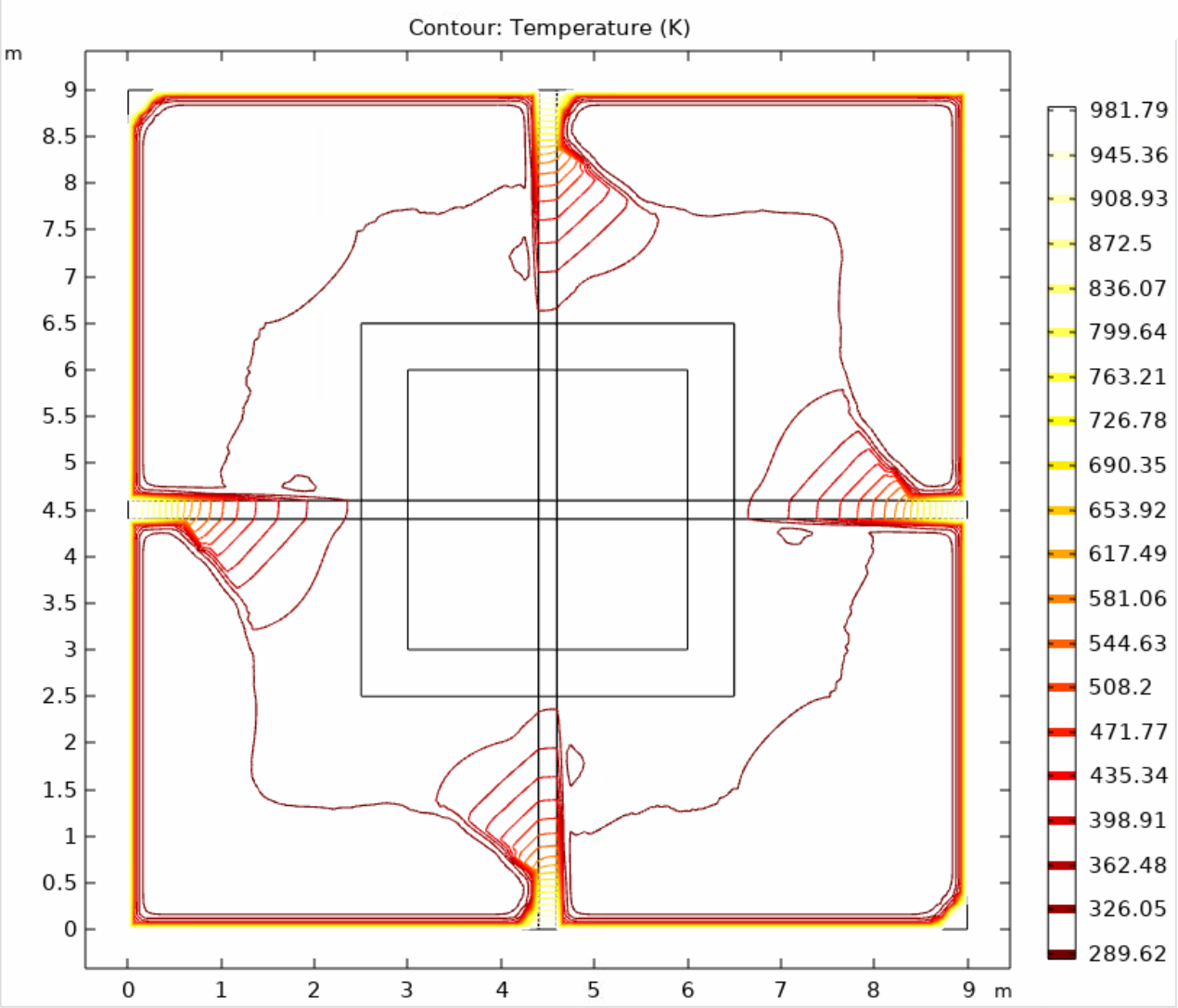}
\caption{Iso-thermal for the response of the optimised system design 2}
\end{figure}

\subsection{Topology Optimisation}
The initial boundary conditions as described in implementation were applied to a design space. A topology optimisation study was added to the COMSOL workspace which utilised the density model as described earlier. The initial response of the non-optimised system was then found as seen in figures 13 and 14 for later comparison. The topology optimisation was run which yielded the topology seen in figure 15. The heat distribution and Iso-thermal contours for this topology can be observed in Figures 16 and 17. Refer to the table found in Parametric Optimisation for the reduction in the objective function. 

Figure 15 shows that the optimiser has determined that the maximum density needed for the material is 0.35. This correlates with a densely packed mesh. This as well as the triangular patterns will be explored further in the analysis and discussion portion of this report. There is a clear distinction between the non-optimised and optimised results. The non-optimised system can be observed to range between $1000 \degree$K to $388.75 \degree$K. This is contrasted greatly with the variance in temperature across the main bulk of the topology for the optimised system which is relatively zero. However, a build-up of heat can be observed on the outer layer of the inside of the vapour chamber. This is much more evident in the plot for the optimised systems iso-thermal contours seen in Figure 17. In an attempt to reduce this build two fins were added to the design space. One standing vertically and one standing horizontally. Figure 15 shows that the bulk of the material for the optimised topology is found roughly 2m from the centre in a square shape. This was also taken into account and a square of solid material was added to the design space 2m from the centre in an attempt to create a design that would be easier to physically create. Figures 18 and 19 show the heat distribution and iso-thermal contours for this design. Figure 20 shows the result of the topology optimisation run over the new design space. Figures 21 and 22 show the heat distribution and iso-thermal contours for the optimised design. It can be observed in Figure 26 that the build-up of heat has reduced from the previous design topology and is not spread more evenly over the surface. However, table 1 shows that the overall temperature gradient over the topology has increased as the objective function has increased meaning that this topology is less optimal than the last. Though this specific topology is slightly less efficient it was shown that the heat build-up can be taken away from the inside surface of the vapour chamber and with a different initial setup perhaps a more efficient topology can be found. 
\section{Analysis  }\label{analysis}

\subsection{Parametric Results}
The conclusions that can be drawn from the parametric results are as follows.

\begin{itemize}
  \item The double fin setup yields a lower objective function and therefore a more optimal setup. This was observed both in the fin number testing and the parametric results. 
  \item Initially the 36 fin setup yielded a lower objective function as shown in the fin number testing. 
  \item Parametric testing showed that the 18-fin setup could yield a lower objective function. This is due to the reduced design space size. Figure 10 shows that increasing the fin size any further than 36 raises the objective function of the mesh. This is because there is not enough space within the squares found between the fins for forced water convection to effectively cool the fins. Figures 11 and 12 show that this causes a build-up of heat on the outside of the inner mesh which ultimately raises the objective function. 
  \item Setup 1C was the most efficient setup from the parametric results. However, figure 15 shows that the solid post element added in to build heat in the centre of the design found a minimum result between the fins meaning that it had no overall effect on the inner mesh. 
   \item There is no conclusive evidence that the solid post of material added into the mesh lowered the objective function as it was thought it would. Table 1 shows the opposite as the objective function was raised from setup 1D to 1C and again from 2D to 2C. 
\end{itemize}

Setup 1C was found to be the most effective in the minimisation of the objective function. This 2D design was then taken from COMSOL and developed into a 3D model for further testing as seen in Figure 27. 

\begin{figure}[h]
\centering
\includegraphics[scale=0.36]{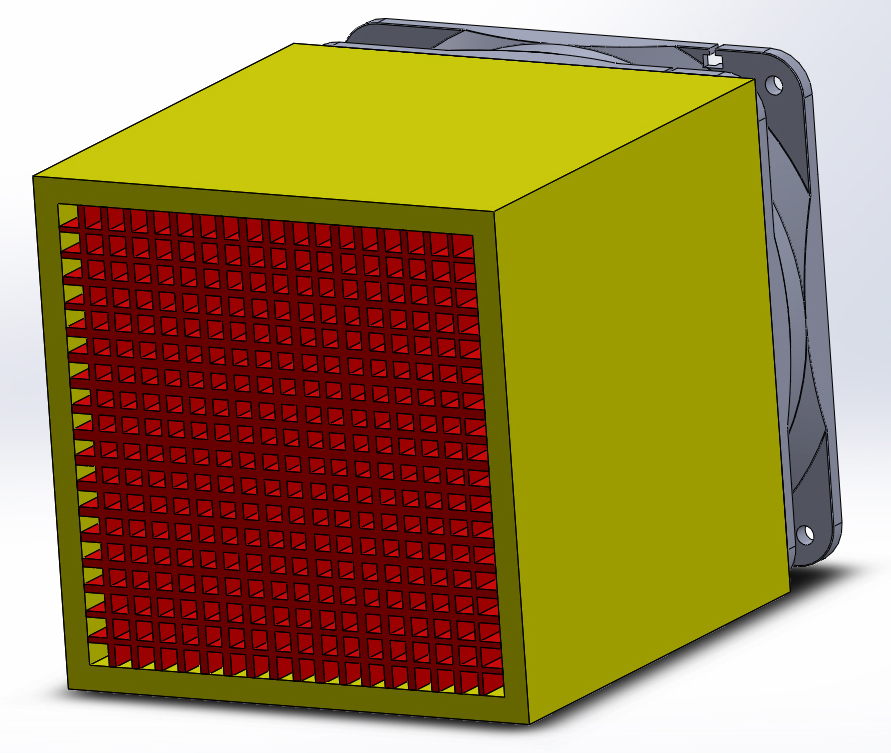}
\caption{3D model for setup 1C for further testing.}
\end{figure}

\subsection{Topology results}
Table 1 shows that of the two density method topology designs tested, topology 1 was more effective at minimising the objective function. Figure 19 shows the density method result which yielded topology 1 from the density method. It shows that there are three distinct triangular meshes with 35 per cent density situated roughly a metre from the inside of the vapour chamber. There are also connections shown at the bottom of the triangles connecting them. Measurements were taken from this result and extrapolated into a feasible geometric design. See Figure 28 below. 

\begin{figure}[h]
\centering
\includegraphics[scale=0.16]{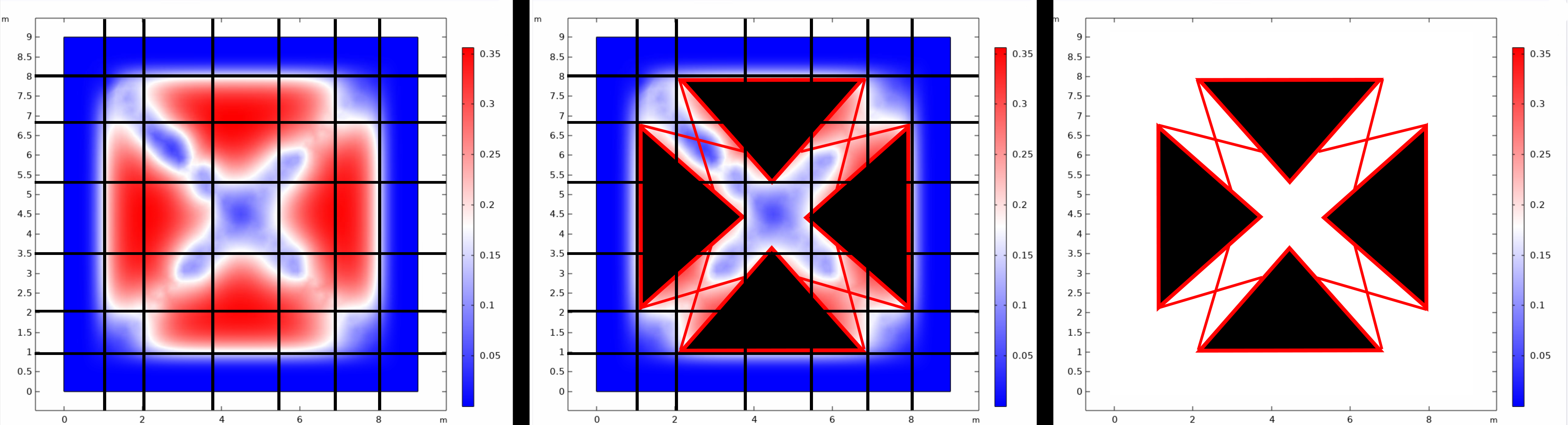}
\caption{Extrapolation of geometry design from the density method topology optimisation results.}
\end{figure}

The objective of this study was to design a feasible heat dissipation system that has application in the industry sector. To meet this requirement the design from the density method topology 1 result was simplified from the haze of 35 percent density mesh coupled with interconnections into an implementable mesh design. It was observed on the parametric results that the double-finned mesh performed the best to dissipate heat. It was also found that the optimal size for these fins was roughly 10cm (see appendix for fin thickness results). From Figure 28 we can see that four symmetric triangles are needed. Their length and height can be read as 5.5m and 2.75m respectively. A triangular mesh design was therefore created fitting these dimensions with 0.1m thick fins spaced 0.28m apart to create a 35 per cent dense design. Two plates were added on either side of the triangular mesh so that they could interface with the interconnects properly. See Figure 29. 

\begin{figure}[h]
\centering
\includegraphics[scale=0.1]{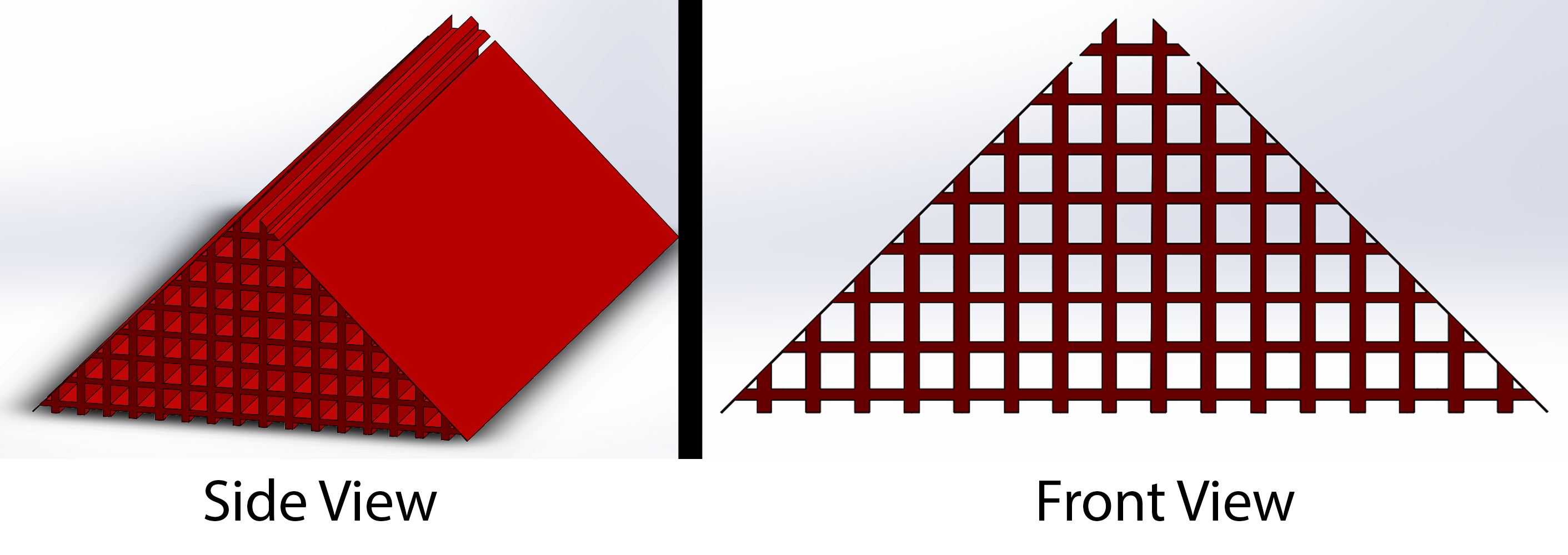}
\caption{Design of triangular mesh from extrapolated geometry.}
\end{figure}

The interconnecting pieces seen at the bottom of each triangle connecting the whole mesh were designed with the same methodology. See Figure 30. 

\begin{figure}[h]
\centering
\includegraphics[scale=0.17]{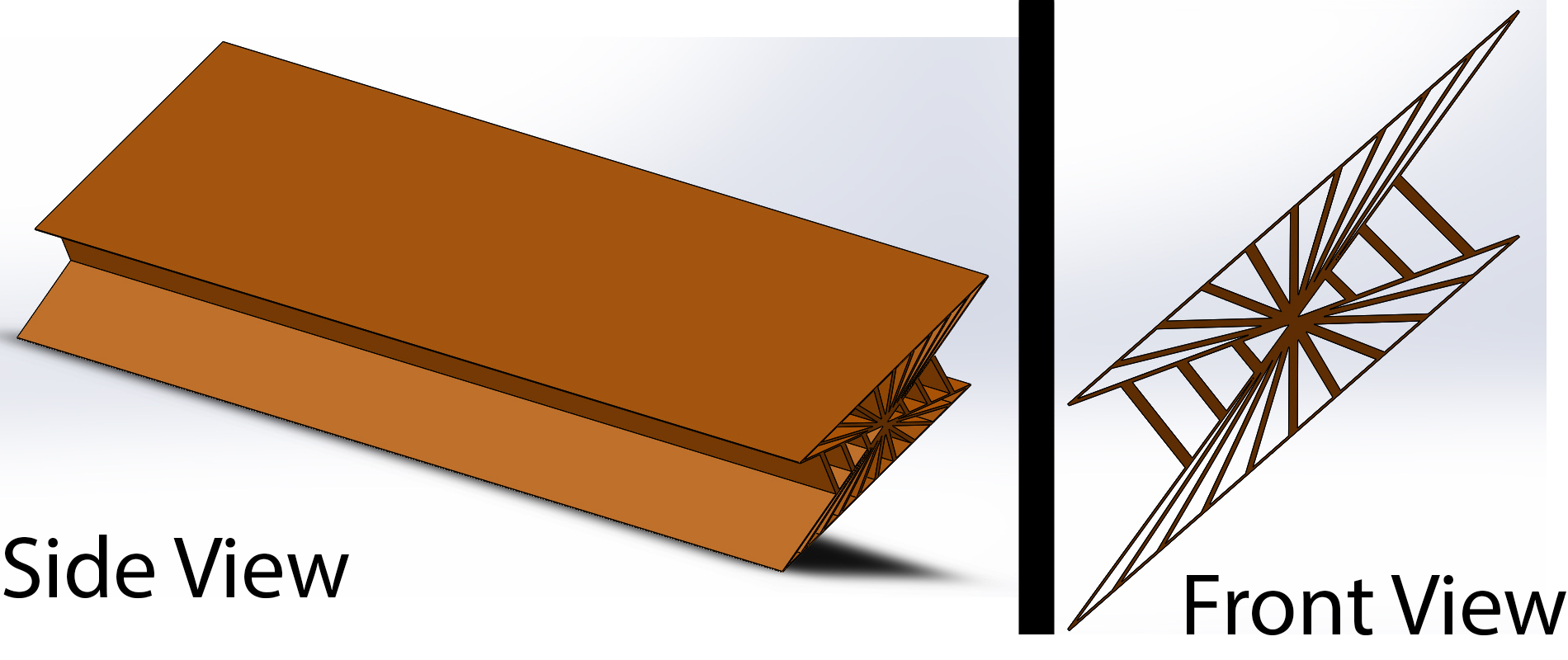}
\caption{Design of interconnecting pieces from extrapolated geometry.}
\end{figure}

These two components were then combined to recreate the topology results yielded from the density method. This mesh was then housed within the vapour chamber from earlier designs. See Figures 31 and 32. 

\begin{figure}[h]
\centering
\includegraphics[scale=0.3]{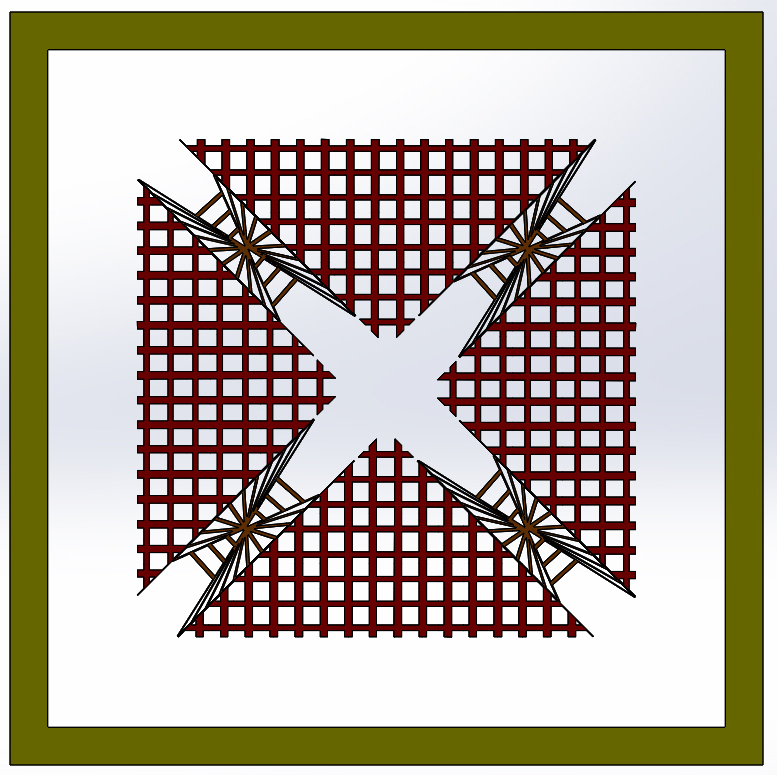}
\caption{Front view of designed heat dissipation system from density method results.}
\end{figure}

\begin{figure}[h]
\centering
\includegraphics[scale=0.3]{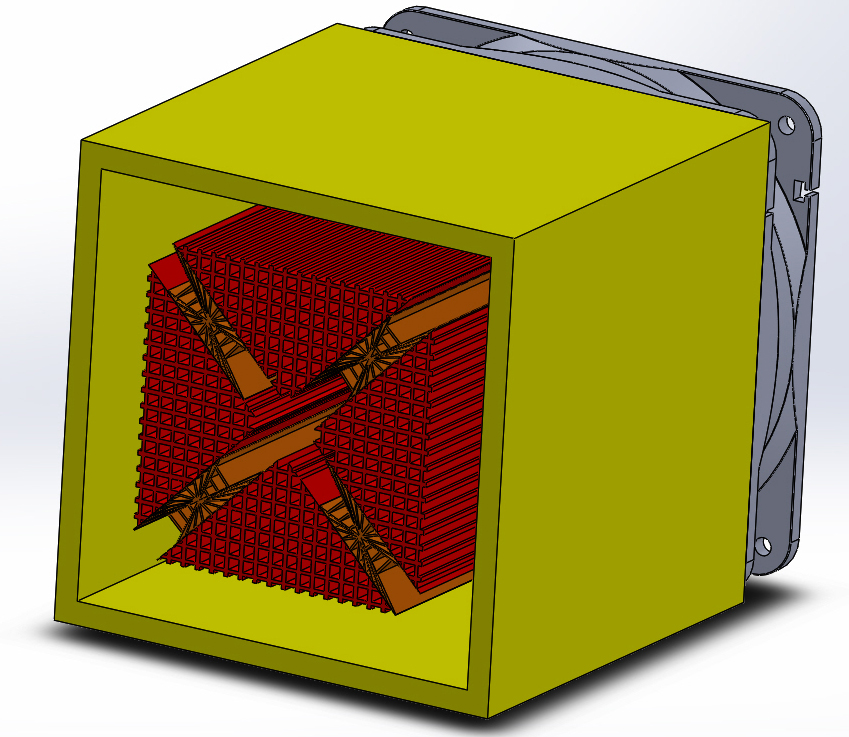}
\caption{side view of designed heat dissipation system from density method results.}
\end{figure}

Both the parametric design and the density method topology design were then implemented and simulated in COMSOL with the boundary conditions as described earlier. The results can be seen in Figures 33 and 34. 

\begin{figure}[h]
\centering
\includegraphics[scale=0.7]{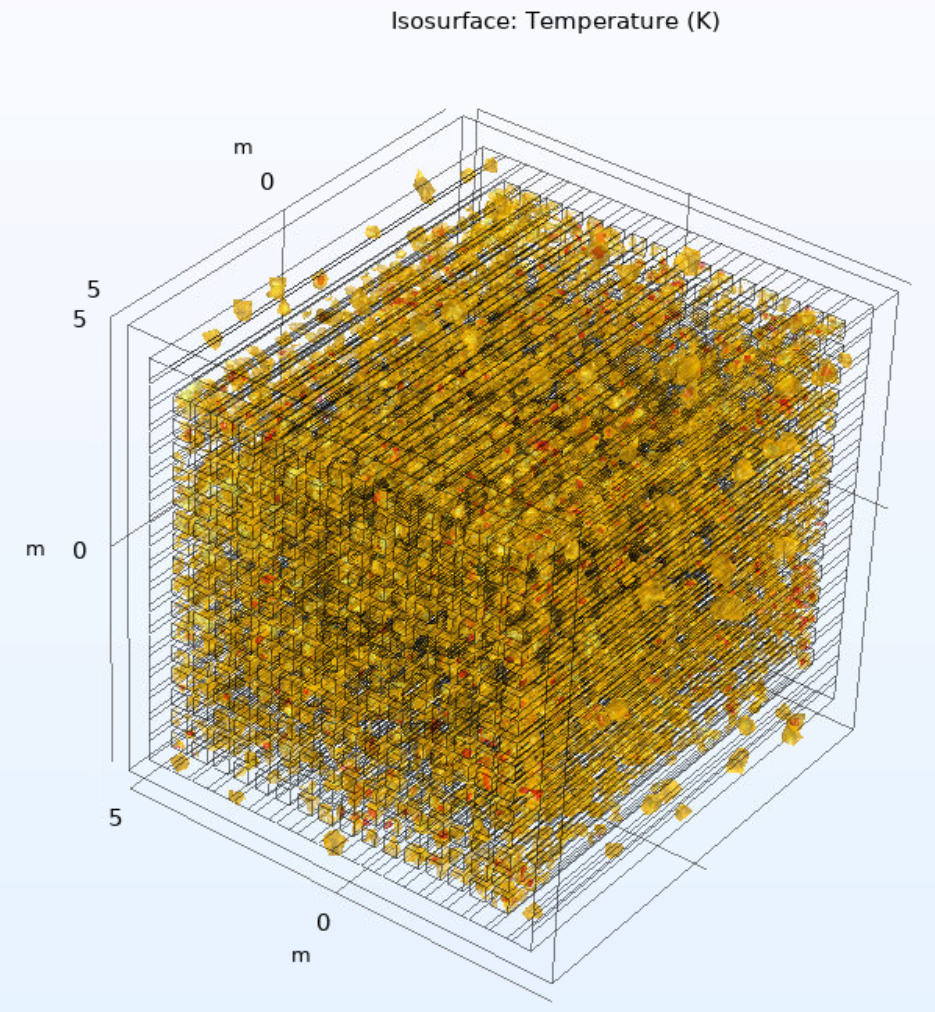}
\caption{Isothermal contour solution for parametric mesh design taken from COMSOL}
\end{figure}

\begin{figure}[h]
\centering
\includegraphics[scale=0.8]{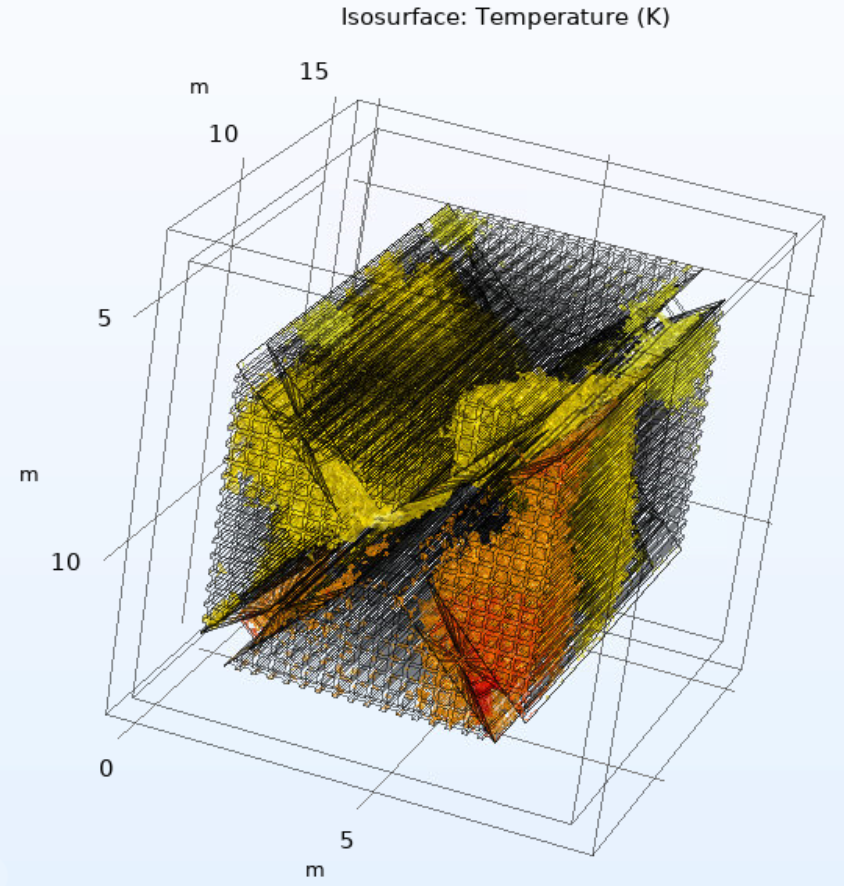}
\caption{Isothermal contour solution for density-based topology optimisation mesh design taken from COMSOL}
\end{figure}

Figure 33 shows the same square iso-thermal contour patterns as was observed in the 2D modelling of the system. It can be observed that there are no major temperature hot spots across the design mesh as was expected. The temperature is distributed evenly across the mesh. Figure 34 shows temperature accumulating in the lower right-hand corner of the mesh. Although there are fewer iso-thermal contours in the second mesh the contours present are larger in magnitude which also correlates well with the 2D simulations of the system. 

\section{Discussion}\label{Discussion}
\subsection{ Comparison of parametric results}
Figures 13 and 15 show the heat distribution for the non-optimised and optimised parametric systems respectively. It can be observed here that through altering the fin thickness even slightly the objective function was lowered from  $843\times 10^{5}$ to  $14.6\times 10^{5}$, a variance of  $828.4\times 10^{5}$ through the slight modification of fin position and thickness. There is a noticeable difference in the temperature gradient. across the two mesh designs. It can be observed that there are fewer cold spots throughout the mesh design which means that the heat is being transported more efficiently to these spots for dissipation through the water domain. Figures 14 and 16 show the iso-thermal contours for the non-optimised and optimised system. Figure 14 shows that there is a noticeable build-up of heat throughout the outside of the mesh design. This is eliminated in Figure 16 showing once again that this mesh more effectively transports its temperature from the outside of the mesh design towards the centre / towards the cold spots of the mesh. 

\subsection{ Comparison of density-based results}
Figure 17 shows the heat distribution for the design space of the density-based method topology optimisation. It can be seen here that the temperature builds on the inner boundaries of the vapour chamber and there is a visible cold spot at the centre of the chamber. Figure 20 shows the heat distribution optimised design space. Close inspection shows that there is an evident build-up of temperature on the inside boundary for the vapour chamber once again. This is most likely due to the time needed for the heat to be transported through the water domain to the inner mesh due to a lack of material seen within the first meter of the inside of the vapour chamber. This build-up of heat is much more evident in the iso-thermal contours of the optimised system shown in Figure 21. This build-up of heat was somewhat eliminated with the secondary design yielded through the density-based topology optimisation method shown in Figures 22-26. The heat distribution in Figure 22 shows that the fins are effectively transporting heat away from the inside boundary of the vapour chamber and Figure 26 shows that the iso-thermal contours are much more spread out over the optimised mesh design. 

\subsection{Comparison of methods}
During this study two separate methods were used to determine the optimal mesh design for a high-temperature heat dissipation system. The objective function used was set to minimise the temperature gradient across the inside of the hollow vapour chamber used so that the forced convection cooling done through a water medium could most effectively cool the whole system. It was found that the density method for topology optimisation yielded the lowest objective function value. However, this method saw a topology designed that allowed the build-up of heat across the inside of the vapour chamber as seen in Figure 21. Figure 20 shows that the inner mesh of the design found through the density method has an almost flat temperature variance. The bulk of of the temperature variance is therefore observed through the build-up of heat on the inside boundary of the vapour chamber. This is not ideal for practical use as it may lead to damage to the system as excessive amounts of heat will be present at its boundary. The second topology design through the density method sought to remedy this situation but did not yield an as efficient design. As such the original topology found through the first set of density method testing was modelled but practically speaking the second topology may have been a better suit for use in industry. The parametric testing saw a mesh design that was almost as efficient as the results from the density method testing but promised to be much more practicable to build and use. Figure 15 shows that the heat distribution over the design is not as evenly spread as was found through the density method but the iso-thermal contours for this design shown in figure 14 show that there is no excessive build-up of heat on any one surface. The selection of which mesh design to use will be a design trade-off between practicality and overall efficiency. Put shortly, the density-based topology optimisation results gave the most efficient mesh for heat dissipation but the parametric optimisation results yielded a mesh design that was comparable in efficiency and is much more practicable in manufacturing and use. 

\subsection{ Application in a system}
Both the designs obtained from parametric optimisation and density-based topology optimisation proved to be effective heat dissipation systems for high-temperature applications. A specific use case for these heat dissipation systems is the generation of electricity from high-temperature sources through the use of a TEG. The two designs were further tested for their heat dissipation properties to observe their suitability in such an application. A time-dependent study was run in COMSOL. A boundary condition for a temperature of  $1000 \degree$K was set to the outside boundary of the vapour chamber. It was found that the system could lower the receiving surface boundary condition to $548.29 \degree$K within three seconds. A high-efficiency TEG uni-couple such as those shown in \cite{TEG_unicouples}, \cite{Teg_unicouples_2}, \cite{compatibility} can be used to yield a FOM in the range of 1.5 \cite{FOM} giving 15 per cent efficiency for temperature to energy conversion. 
Using a hot side temperature of  $1000 \degree$K and a cold side temperature of  $548.29 \degree$K we can see that through equation 1 an overall efficiency of 12.4 per cent can be achieved for a TEG unit.  Utilising pulse mode operation can further improve the power output by 2.7x \cite{pulse_mode} meaning an efficiency of 33 per cent is possible for a TEG unit.

\subsection{Application in Industry}
The general efficiency of a coal-fired power station ranges between ~30-50\% with most of the losses occurring as heat within the range of $600 - 1200\degree$K \cite{coal}. Miscibility Gap Alloys (MGA's) is a new technology that has been developed by the University of Newcastle for the storage of heat energy. An energy density of 1.84 has been reported for MGA technology at a temperature of  $1358.15\degree$K \cite{MGA} which is comparable to most modern-day Lithium Ion batteries. A specific use case for an application of this technology in industry would be the recapture of waste heat from coal-fired power stations for storage in MGA blocks for later use. The heat stored within these MGA blocks can then be used with a high FOM TEG-Unicouple utilising a highly efficient heat dissipation system such as was developed through this study. This system could further use the waste heat rejected from the heat dissipation system to power secondary TEGs. The power generated from these secondary TEGs could then be used in a control feedback system to run a motor and allow the system to operate at pulse mode intervals. 

\section{Conclusion}\label{conclusion}
In this project, two water-cooled forced convection heat dissipation systems were developed for high-temperature applications. The design was done using both parametric optimisation and density-based topology optimisation. Two optimal mesh designs were produced using these separate methods with the variance in the objective function between them minimal. It was found that for the parametric case, a double-finned setup utilising 18 fins worked best. For the topology-based optimisation, it was found that the optimal solution placed a 35 per cent dense mesh within the inside of the design space. This mesh was then realised using some of the information gained from the parametric testing. COMSOL was then used to test these systems and it was found that, with the use of pulse mode operation a 33 percent efficient TEG unit was plausible. A use case was then proposed which saw the use of MGA units to store waste heat from coal-fired power stations for later use in power generation through the TEG system utilising the optimal heat dissipation systems.

\bibliographystyle{IEEEtran}
\bibliography{references}

\begin{thebibliography}{10}
\providecommand{\url}[1]{#1}
\csname url@rmstyle\endcsname
\providecommand{\newblock}{\relax}
\providecommand{\bibinfo}[2]{#2}
\providecommand\BIBentrySTDinterwordspacing{\spaceskip=0pt\relax}
\providecommand\BIBentryALTinterwordstretchfactor{4}
\providecommand\BIBentryALTinterwordspacing{\spaceskip=\fontdimen2\font plus
\BIBentryALTinterwordstretchfactor\fontdimen3\font minus
  \fontdimen4\font\relax}
\providecommand\BIBforeignlanguage[2]{{%
\expandafter\ifx\csname l@#1\endcsname\relax
\typeout{** WARNING: IEEEtran.bst: No hyphenation pattern has been}%
\typeout{** loaded for the language `#1'. Using the pattern for}%
\typeout{** the default language instead.}%
\else
\language=\csname l@#1\endcsname
\fi
#2}}

\bibitem{FOM}
\BIBentryALTinterwordspacing
D.~Aswal, R.~Basu, and A.~Singh, ``Key issues in the development of
  thermoelectric power generators: High figure-of-merit materials and their
  highly conducting interfaces with metallic interconnects,'' \emph{Technical
  Physics Division, Bhabha Atomic Research Center}, 2019, accessed: 08 June
  2020. [Online]. Available:
  \url{https://www.sciencedirect.com/science/article/pii/S0196890416300036}
\BIBentrySTDinterwordspacing

\bibitem{coal}
\BIBentryALTinterwordspacing
M.~Suresh, K.~Reddy, and A.~Kolar, ``Thermodynamic analysis of a coal-fired
  power plant repowered with pressurized pulverized coal combustion,''
  \emph{Institute of Mechanical Engineers}, 2010, accessed: 08 June 2020.
  [Online]. Available: \url{https://doi.org/10.1177/0957650911418421}
\BIBentrySTDinterwordspacing

\bibitem{exergy}
\BIBentryALTinterwordspacing
J.~Cullen and J.~Allwood, ``The efficient use of energy: Tracing the global
  flow of energy from fuel to service,'' \emph{Department of Engineering,
  University of Cambridge, Cambridge}, 2010, accessed: 08 June 2020. [Online].
  Available: \url{https://doi.org/10.1016/j.enpol.2009.08.054}
\BIBentrySTDinterwordspacing

\bibitem{Heat_sinks_TEG}
\BIBentryALTinterwordspacing
A.~Elghool, F.~Basrawi, T.~Ibrahim, K.~Habib, H.~Ibrahim, and D.~Idris, ``A
  review on heat sink for thermo-electric power generation: Classifications and
  parameters affecting performance,'' \emph{Energy Sustainability Focus Group
  (ESFG), Faculty of Mechanical Engineering, Universiti Malaysia Pahang}, 2017,
  accessed: 08 June 2020. [Online]. Available:
  \url{https://www.sciencedirect.com/science/article/pii/S0196890416311414}
\BIBentrySTDinterwordspacing

\bibitem{Vapour_chamber1}
\BIBentryALTinterwordspacing
S.~Wiriyasart and P.~Naphon, ``Thermal management system with different
  configuration liquid vapor chambers for high power electronic devices,''
  \emph{Department of Mechanical Engineering, Faculty of Engineering,
  Srinakharinwirot University}, 2019, accessed: 08 June 2020. [Online].
  Available:
  \url{https://www.sciencedirect.com/science/article/pii/S2214157X1930512X}
\BIBentrySTDinterwordspacing

\bibitem{Vapour_chamber2}
\BIBentryALTinterwordspacing
M.~Bulut, S.~Kandlikar, and N.~Sozbir, ``A review of vapor chambers,''
  \emph{Heat Transfer Engineering}, vol.~40, no.~19, pp. 1551--1573, 2019,
  accessed: 08 June 2020. [Online]. Available:
  \url{https://doi.org/10.1080/01457632.2018.1480868}
\BIBentrySTDinterwordspacing

\bibitem{Lange_paper}
\BIBentryALTinterwordspacing
F.~Lange, C.~Hein, G.~Li, and C.~Emmelmann, ``Numerical optimization of active
  heat sinks considering restrictions of selective laser melting,''
  \emph{Fraunhofer Research Institution for Additive Manufacturing Technologies
  IAPT, Hamburg, Germany; Institute of Laser and System Technologies, Hamburg
  University of Technology, Germany}, 2018, accessed: 08 June 2020. [Online].
  Available:
  \url{https://www.researchgate.net/publication/328738213_Numerical_optimization_of_active_heat_sinks_considering_restrictions_of_selective_laser_melting}
\BIBentrySTDinterwordspacing

\bibitem{Topology_Webpage}
\BIBentryALTinterwordspacing
K.~Jensen, ``Performing topology optimization with the density method,'' COMSOL
  Blog, 2019, accessed: 08 June 2020. [Online]. Available:
  \url{https://www.comsol.com/blogs/performing-topology-optimization-with-the-density-method/}
\BIBentrySTDinterwordspacing

\bibitem{TEG_unicouples}
\BIBentryALTinterwordspacing
H.~M. El-Genk and H.~Saber, ``High efficiency segmented thermoelectric
  unicouples for operation between 973 and 300 k,'' \emph{Department of
  Chemical and Nuclear Engineering, Institute for Space and Nuclear Power
  Studies, The University of New Mexico}, 2003, accessed: 08 June 2020.
  [Online]. Available:
  \url{https://www.sciencedirect.com/science/article/pii/S0196890402001097}
\BIBentrySTDinterwordspacing

\bibitem{Teg_unicouples_2}
\BIBentryALTinterwordspacing
T.~Caillat, A.~Borshchevsky, J.~Snyder, and J.~Fleurial, ``High efficiency
  segmented thermoelectric unicouples,'' in \emph{AIP Conference Proceedings},
  2001, accessed: 08 June 2020. [Online]. Available:
  \url{https://aip.scitation.org/doi/10.1063/1.1358058}
\BIBentrySTDinterwordspacing

\bibitem{compatibility}
\BIBentryALTinterwordspacing
J.~Snyder, ``Application of the compatibility factor to the design of segmented
  and cascaded thermoelectric generators,'' 2004, accessed: 08 June 2020.
  [Online]. Available: \url{https://aip.scitation.org/doi/10.1063/1.1689396}
\BIBentrySTDinterwordspacing

\bibitem{pulse_mode}
\BIBentryALTinterwordspacing
M.~Haras, M.~Markiewicz, M.~Monfray, and T.~Skotnicki, ``Pulse mode of
  operation – a new booster of teg, improving power up to 2.7 times – to
  better fit iot requirements,'' \emph{Warsaw University of Technology,
  Poland}, 2019, accessed: 08 June 2020. [Online]. Available:
  \url{https://www.sciencedirect.com/science/article/pii/S2211285519309115}
\BIBentrySTDinterwordspacing

\bibitem{MGA}
\BIBentryALTinterwordspacing
D.~Cuskelly, B.~Fraser, S.~Reed, A.~Post, M.~Copus, and E.~Kisi, ``Thermal
  storage for csp with miscibility gap alloys,'' \emph{The University of
  Newcastle, NSW, Australia}, 2012, accessed: 08 June 2020. [Online].
  Available: \url{https://doi.org/10.1016/j.applthermaleng.2012.11.029}
\BIBentrySTDinterwordspacing

\end{thebibliography}


\begin{thebibliography}{9}

\bibitem{FOM} 
Aswal, D. Basu, R. Singh, A. (2019).
\textit{Key issues in the development of thermoelectric power generators: High ﬁgure-of-merit materials and their highly conducting interfaces with metallic interconnects}.Technical Physics Division, Bhabha Atomic Research Center, Mumbai.  Available at:  https://www.sciencedirect.com/science/article/pii/S0196890416300036 (Accessed: 08 June 2020).


\bibitem{Lange paper}
Lange, F. Hein, c.Li, G. Emmelmann C. (2018).
\textit{Numerical optimization of active heat sinks considering restrictions
of selective laser melting
}. 1. Fraunhofer Research Institution for Additive Manufacturing Technologies IAPT, Hamburg,
Germany
2. Institute of Laser and System Technologies, Hamburg University of Technology, Germany Available at:  https://www.researchgate.net/publication/328738213\_Numerical
\_optimization\_of\_active\_heat\_sinks\_considering\_restrictions\_of\_selective
\_laser\_melting (Accessed: 08 June 2020).

\bibitem{Topology Webpage} 
Jensen, K. (2019).
\textit{Performing Topology Optimization with the Density Method}. COMSOL Blog. Available at:  https://www.comsol.com/blogs/performing-topology-optimization-with-the-density-method/ (Accessed: 08 June 2020).

\bibitem{Hartel} 
Jan H. K. Haertel, Kurt Engelbrecht, Boyan S. Lazarov, Ole Sigmund (2018).
\textit{Topology optimization of a pseudo 3D thermofluid heat sink model}. International Journal of Heat and Mass Transfer 121. Available at:  https://www.researchgate.net/publication/323869595\_Topology\_
optimization\_of\_a\_pseudo\_3D\_thermofluid\_heat\_sink\_model (Accessed: 08 June 2020).


\bibitem{Hartel_2} 
Jan H. K. Haertel, Tian Lei, Joe Alexandersen, Kurt Engelbrecht, Boyan S. Lazarov, Ole Sigmund (2017).
\textit{Performing Topology Optimization with the Density Method}. Technical University of Denmark. Available at:  https://backend.orbit.dtu.dk/ws/portalfiles/portal/138508858/
DanishDays2017\_Abstract\_Haertel.pdf (Accessed: 08 June 2020).

\bibitem{Heat_sinks_TEG} 
Elghool, A. Basrawi F. ,Ibrahim, T. Habib, K. Ibrahim, H. Idris,D. (2017).
\textit{A review on heat sink for thermo-electric power generation: Classiﬁcations and parameters affecting performance}.Energy Sustainability Focus Group (ESFG), Faculty of Mechanical Engineering, Universiti Malaysia Pahang.  Available at:  https://www.sciencedirect.com/science/article/pii/S0196890416311414 (Accessed: 08 June 2020).

\bibitem{TEG_improve_2_stage} 
Wang, C. Hung, C. Chen, W. (2012).
\textit{Design of heat sink for improving the performance of thermoelectric generator using two-stage optimization}.  Department of Mechanical Engineering, National Cheng Kung University, Tainan 701, Taiwan.  
Available at:  https://www.sciencedirect.com/science/article/pii/S0360544212000308 (Accessed: 08 June 2020).

\bibitem{two_phased_natural_circulation} 
Wang, D.Jiang, J. Zhan, D. Zhang, X. Liu, X. (2019).
\textit{Design and analysis of improved two-phase natural circulation systems with thermoelectric generator}.Sino-French Institute of Nuclear Engineering and Technology, Sun Yet-Sen University, Zhuhai, Guangdong.  Available at: https://www.sciencedirect.com/science/article/pii/S030645491930790X (Accessed: 08 June 2020).

\bibitem{Exhaust} 
F.P. Britoa N. Pachecoa, R. Vieiraa, J. Martinsa, L. Martinsa, J. Teixeiraa, L.M. Goncalvesd, (2019).
\textit{Efficiency improvement of vehicles using temperature controlled exhaust thermoelectric generators}. University of Minho. Available at:  https://www.sciencedirect.com/science/article/pii/S0196890419312610 (Accessed: 08 June 2020).

\bibitem{TEG_Heat_Transfer_pipes} 
Yang, Y. Wang, S. Zhu, Y. (2020).
\textit{Evaluation method for assessing heat transfer enhancement effect on performance improvement of thermoelectric generator systems}. School of Mechanical Engineering, Tianjin University, PR China.  Available at:  https://www.sciencedirect.com/science/article/pii/S0306261920302002 (Accessed: 08 June 2020).


\bibitem{TEG_TEC} 
Kwan, T. Wu, X. Yao, Q. (2019).
\textit{Complete implementation of the combined TEG-TEC temperature control and energy harvesting system}. School of Aeronautics and Astronautics, Sun Yat-Sen University, China. School of AMME, University of Sydney, Sydney, Australia.  https://www.sciencedirect.com/science/article/pii/S0967066119301935 (Accessed: 08 June 2020).

\bibitem{exhaust_teg2} 
A. Massaguerb, T. Pujola, M. Comamalaa, E. Massaguera (2019).
\textit{Feasibility study on a vehicular thermoelectric generator coupled to an exhaust gas heater to improve aftertreatment’s efficiency in cold-starts}. Department of Mechanical Engineering and Industrial Construction, University of Girona, C. Universitat de Girona, 4, 17003 Girona, Spain Available at:  https://www.sciencedirect.com/science/article/pii/S1359431119339870 (Accessed: 08 June 2020).

\bibitem{pulse_mode}
Haras, M., Markiewicz, M., Monfray, M., Skotnicki, T. (2019).
\textit{Pulse mode of operation – A new booster of TEG, improving power up to 2.7 times – to better fit IoT requirements}.
Warsaw University of Technology, Poland. 
Available at: \url{https://www.sciencedirect.com/science/article/pii/S2211285519309115}.
Accessed: 08 June 2020.


\bibitem{compatibility} 
Snyder, J (2004).
\textit{Application of the compatibility factor to the design of segmented and cascaded thermoelectric generators}. Available at:  https://aip.scitation.org/doi/10.1063/1.1689396 (Accessed: 08 June 2020).


\bibitem{topology_optimisation-2} 
Santhanakrishnan, M.Tilford, T. Bailey, C. (2017).
\textit{Performing Topology Optimization with the Density Method}. Department of Mathematical Sciences, The University of Greenwich, London, UK. Available at:  https://www.comsol.com/paper/level-set-based-topology-optimisation-of-convectively-cooled-heatsinks-52121 (Accessed: 08 June 2020).

\bibitem{heat_sink_optimisation} 
 Stadler , M. (2019).
\textit{OPTIMIZATION OF THE GEOMETRY OF A HEAT SINK}. University of Virginia, Charlottesville, VA. 
Available at:  https://www.mech.kth.se/~matds/docs/mbsvsgcheatsinkopt.pdf (Accessed: 08 June 2020).

\bibitem{Solar_TEG} 
Jena S. et al. (2020)
\textit{Power Generation from Various Interconnecting Solar PV Networks for an Electrically Coupled Solar PV-TEG System Under Healthy and Partly Cloudy Condition}. In: Sharma R., Mishra M., Nayak J., Naik B., Pelusi D. (eds) Innovation in Electrical Power Engineering, Communication, and Computing Technology. Lecture Notes in Electrical Engineering, vol 630. Springer, Singapore. Available at:  https://link.springer.com/chapter/10.1007/978-981-15-2305-2\_39 (Accessed: 08 June 2020).

\bibitem{TEG_unicouples} 
El-GenkHamed, M. Saber, H. (2003).
\textit{High efficiency segmented thermoelectric unicouple for operation between 973 and 300 K}. Department of Chemical and Nuclear Engineering, Institute for Space and Nuclear Power Studies,The University of New Mexico. Available at:  https://www.sciencedirect.com/science/article/pii/S0196890402001097 (Accessed: 08 June 2020).

\bibitem{TEG_unicouples_2} 
 Caillat, T. Borshchevsky, A. Snyder, J. Fleurial. JP. (2001).
\textit{High efficiency segmented thermoelectric unicouples}. AIP Conference Proceedings. Available at:  https://aip.scitation.org/doi/10.1063/1.1358058 (Accessed: 08 June 2020).


\bibitem{Vapour_chamber1} 
Wiriyasart, S. Naphon, P. (2019).
\textit{Thermal management system with different configuration liquid vapor chambers for high power electronic devices}. Department of Mechanical Engineering, Faculty of Engineering, Srinakharinwirot University.  Available at:  https://www.sciencedirect.com/science/article/pii/S2214157X1930512X (Accessed: 08 June 2020).

\bibitem{Vapour_chamber2} 
Bulut, M. Kandlikar, S. Sozbir, N. (2019)
\textit{A Review of Vapor Chambers}. Heat Transfer Engineering, 40:19, 1551-1573. Available at:  https://doi.org/10.1080/01457632.2018.1480868 (Accessed: 08 June 2020).


\bibitem{exergy} 
Cullen, J. Allwood, J. (2010)
\textit{The efficient use of energy: Tracing the global flow of energy from fuel to service}. Department of Engineering, University of Cambridge, Cambridge. Available at:  https://doi.org/10.1016/j.enpol.2009.08.054 (Accessed: 08 June 2020).


\bibitem{coal} 
Suresh, M. Reddy, K. Kolar, A. (2010)
\textit{Thermodynamic analysis of a coal-fired power plant repowered with pressurized pulverized coal combustion}. Institute of Mechanical Engineers. Available at:  https://doi.org/10.1177/0957650911418421 (Accessed: 08 June 2020).

\bibitem{MGA} 
Cuskelly, D. Fraser,B. Reed, S. Post, A. Copus, M. Kisi, E. (20120)
\textit{Thermal Storage for CSP with Miscibility Gap Alloys}.  The University of Newcastle, NSW, Australia . Available at:  https://doi.org/10.1016/j.applthermaleng.2012.11.029. (Accessed: 08 June 2020).


\end{thebibliography}

\section{Appendix}\label{appendix}

\makenomenclature

\mbox{}

\nomenclature{$\eta$}{Efficiency of a TEG}
\nomenclature{$T_H$}{Temperature at the hot side of the TEG}
\nomenclature{$T_L$}{Temperature at the cold side of the TEG}
\nomenclature{$ZT_H$}{FOM of TEG at high temperature}
\nomenclature{$ZT_M$}{FOM of TEG at average temperature}
\nomenclature{$ZT_{thermoelement} $}{FOM for an individual thermoelement from TEG device}
\nomenclature{$\alpha$}{seebeck coefficient}
\nomenclature{$\sigma$}{electrical conductivity}
\nomenclature{$k$}{Thermal conductivity}
\nomenclature{$k_e$}{Electronic thermal conductivity}
\nomenclature{$k_{Bi}$}{Bipolar thermal conductivity}
\nomenclature{$k_L$}{Lattice thermal conductivity}
\nomenclature{$T$}{Temperature state variable}
\nomenclature{$Q$}{Volumetric heat generation}
\nomenclature{$\rho$}{Fluid density}
\nomenclature{$\eta_f$}{Fluid dynamic viscosity}
\nomenclature{$P$}{Fluid pressure state variable vector}
\nomenclature{$U$}{Velocity field state variable vector}
\nomenclature{$\alpha_{por}$}{ inverse permeability of porous medium}
\nomenclature{$C$}{Heat capacity state variable vector}
\nomenclature{$f_{obj(x)}$}{Objective function}
\nomenclature{$f_{con}$}{Constraint functions}
\nomenclature{$\Omega$}{Design space / Design variables vector}
\nomenclature{t}{fin thickness}
\nomenclature{H}{Solid post height}
\nomenclature{W}{Solid post width}
\nomenclature{A}{Solid post thickness - x}
\nomenclature{B}{Solid post thickness - y}
\nomenclature{q}{Linear combination parameter}
\nomenclature{$h_0$}{Size of details in solution}
\nomenclature{$h_{max}$}{Mesh size}
\nomenclature{$\theta_c$}{material density}
\nomenclature{$\theta_f$}{Minimum material density}
\nomenclature{$R_{min}$}{minimum Helmholtz filter radius}
\nomenclature{$k_{SIMP}$}{Thermal SIMP penalization factor}
\nomenclature{$C_{SIMP}$}{Specific heat SIMP penalization factor}
\nomenclature{$\theta_p$}{Interpolation SIMP exponent}
\nomenclature{$\rho_{des}$}{Design space density}
\printnomenclature

\clearpage

\subsection{Parametric size results}

\begin{figure}[h]
\centering
\includegraphics[scale=0.6]{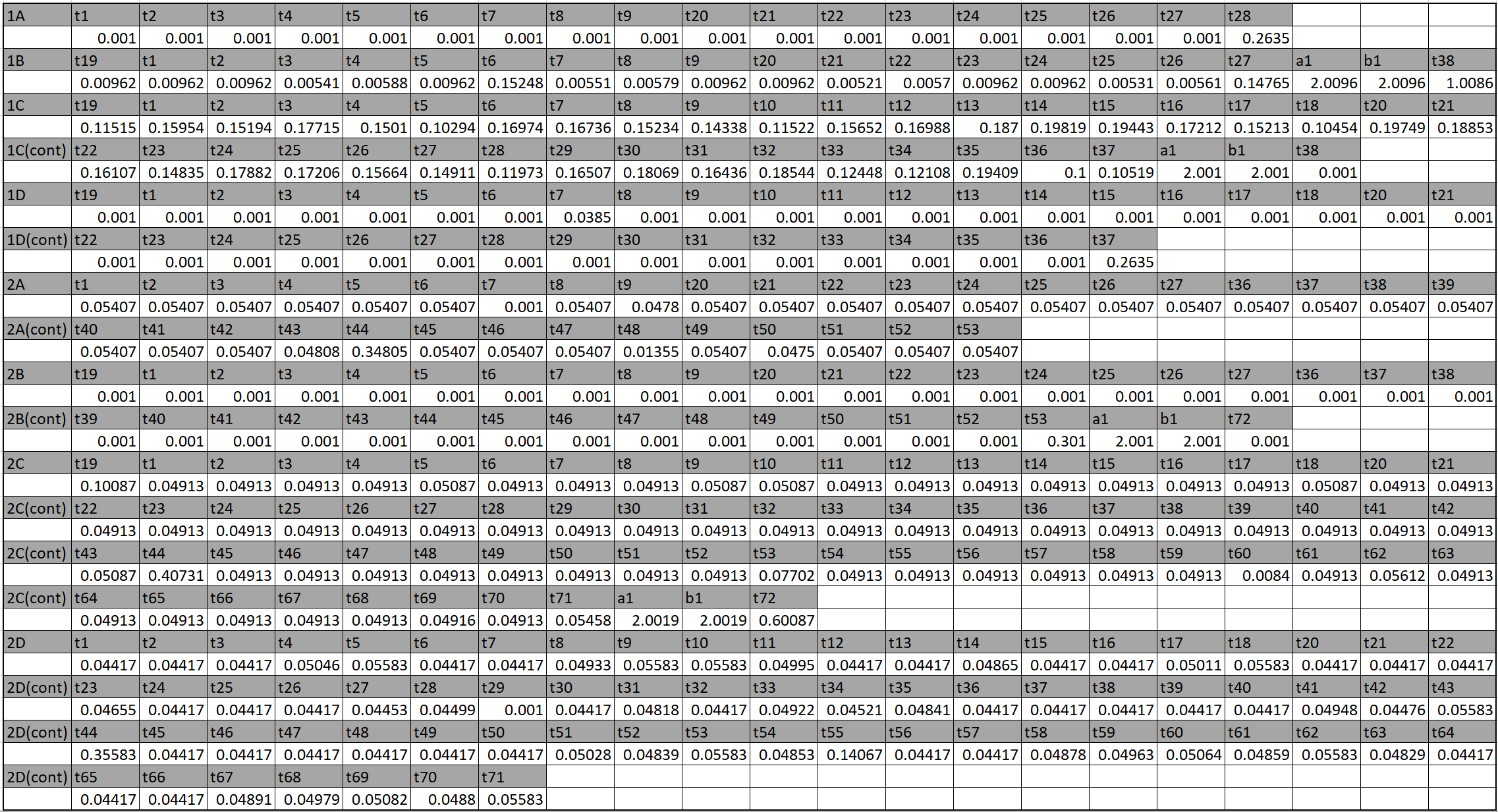}
\caption{Fin thickness, post thickness width and height results from parametric testing }
\end{figure}

\clearpage
\subsection{ Further images of designs}

\begin{figure}[h]
\centering
\includegraphics[scale=0.3]{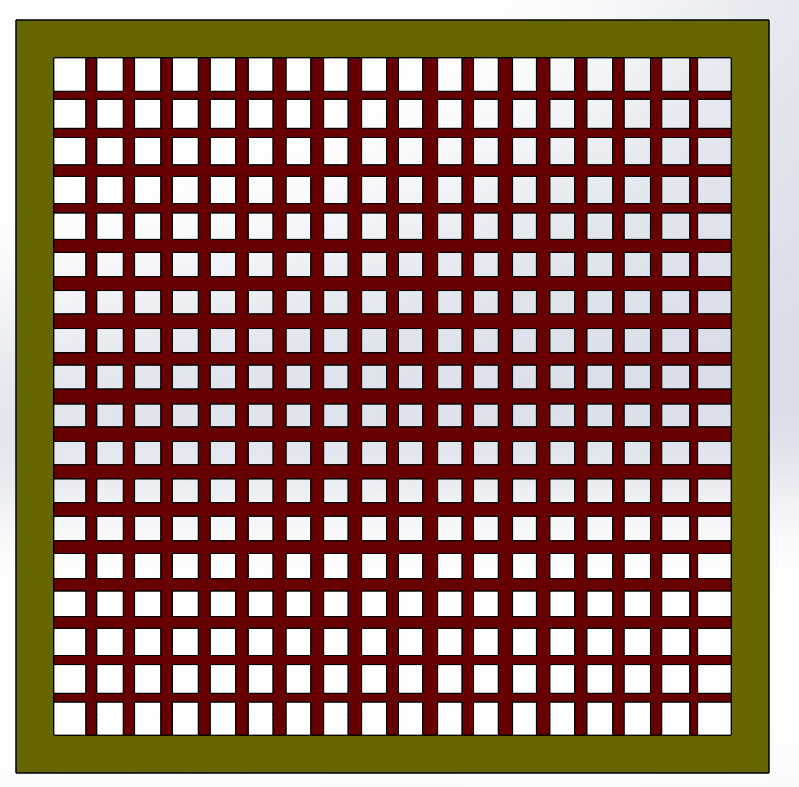}
\caption{Parametric}
\end{figure}

\begin{figure}[h]
\centering
\includegraphics[scale=0.3]{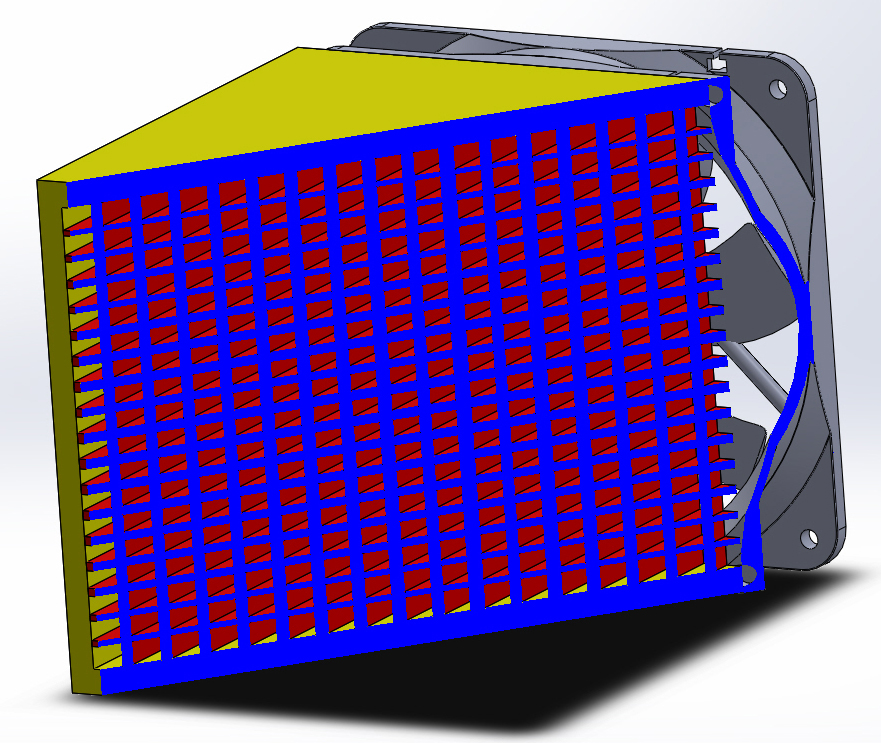}
\caption{Parametric}
\end{figure}

\begin{figure}[h]
\centering
\includegraphics[scale=0.3]{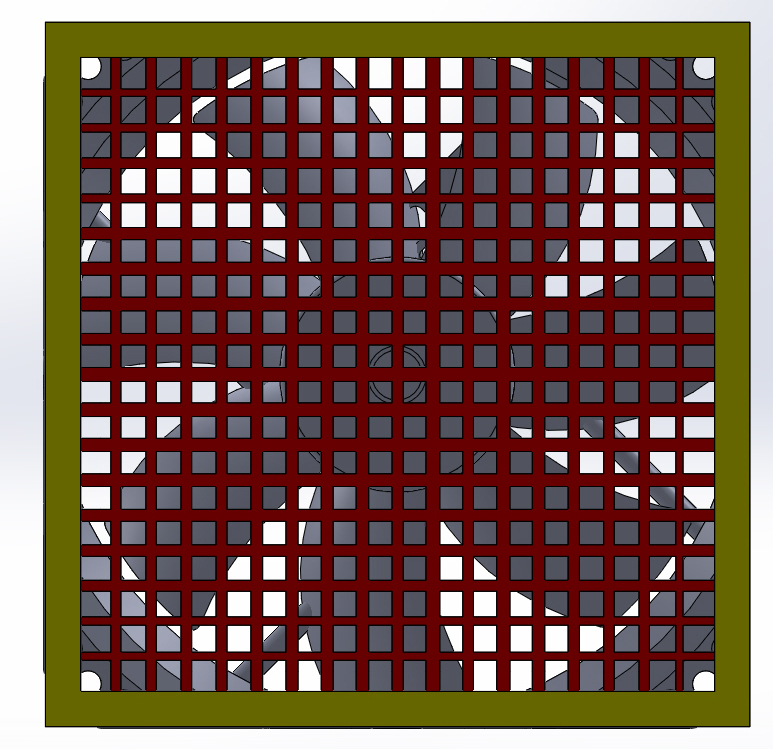}
\caption{Parametric}
\end{figure}

\begin{figure}[h]
\centering
\includegraphics[scale=0.3]{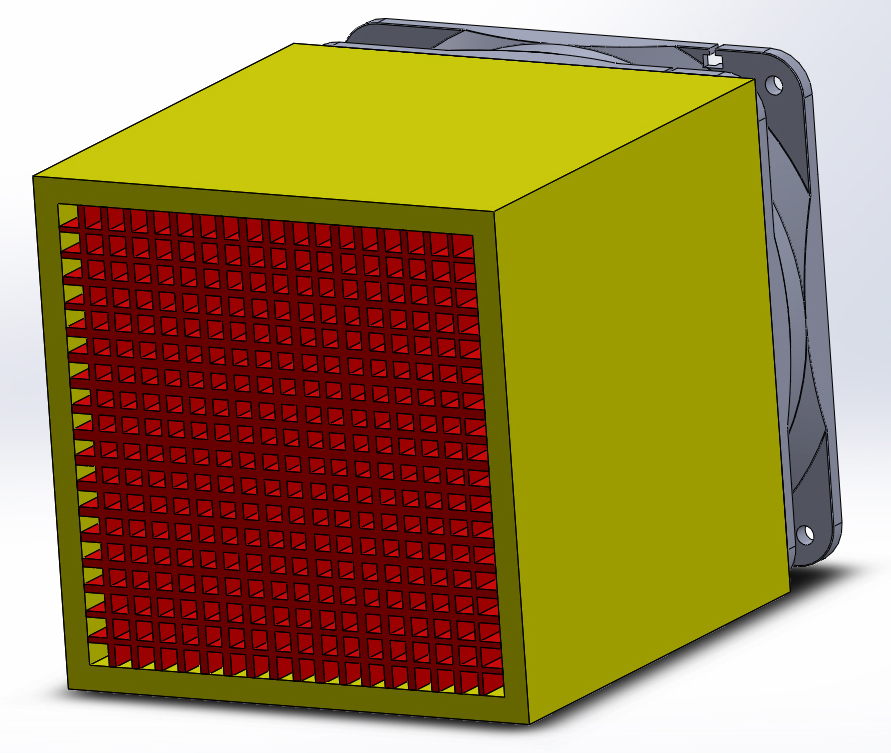}
\caption{Parametric}
\end{figure}

\clearpage

\begin{figure}[h]
\centering
\includegraphics[scale=0.3]{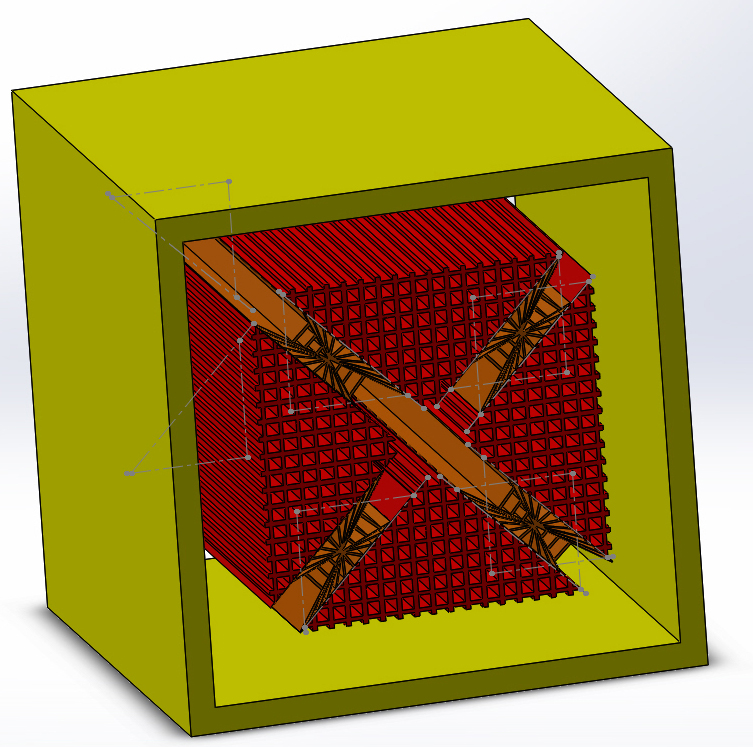}
\caption{Density based topology}
\end{figure}

\begin{figure}[h]
\centering
\includegraphics[scale=0.3]{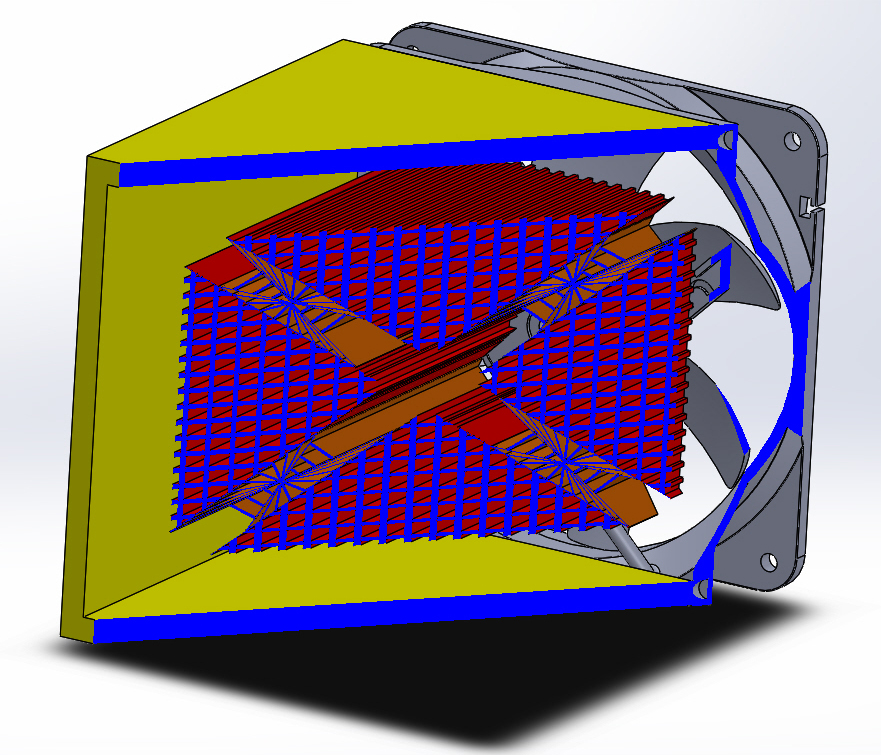}
\caption{Density based topology}
\end{figure}

\begin{figure}[h]
\centering
\includegraphics[scale=0.3]{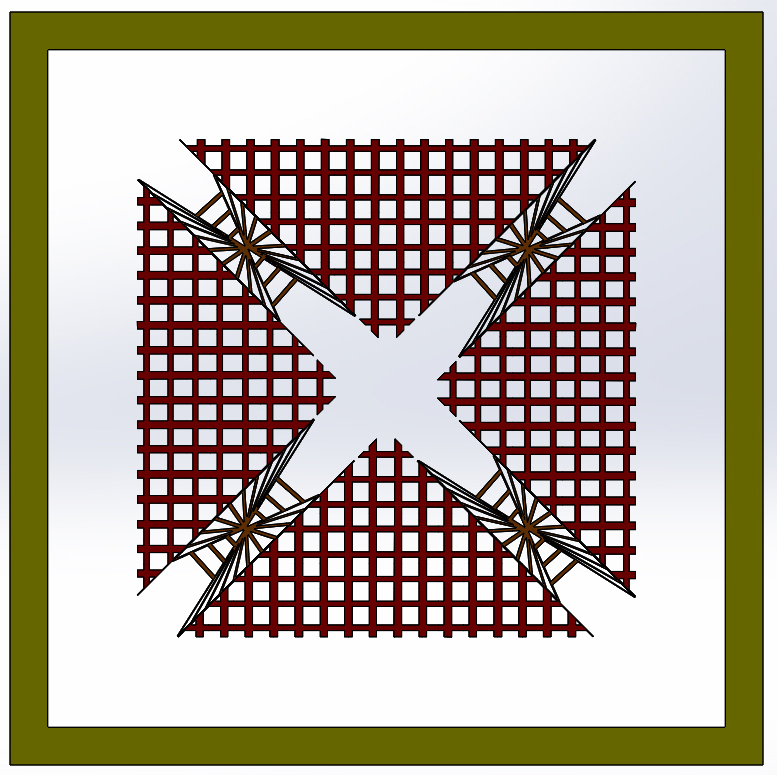}
\caption{Density based topology}
\end{figure}

\begin{figure}[h]
\centering
\includegraphics[scale=0.3]{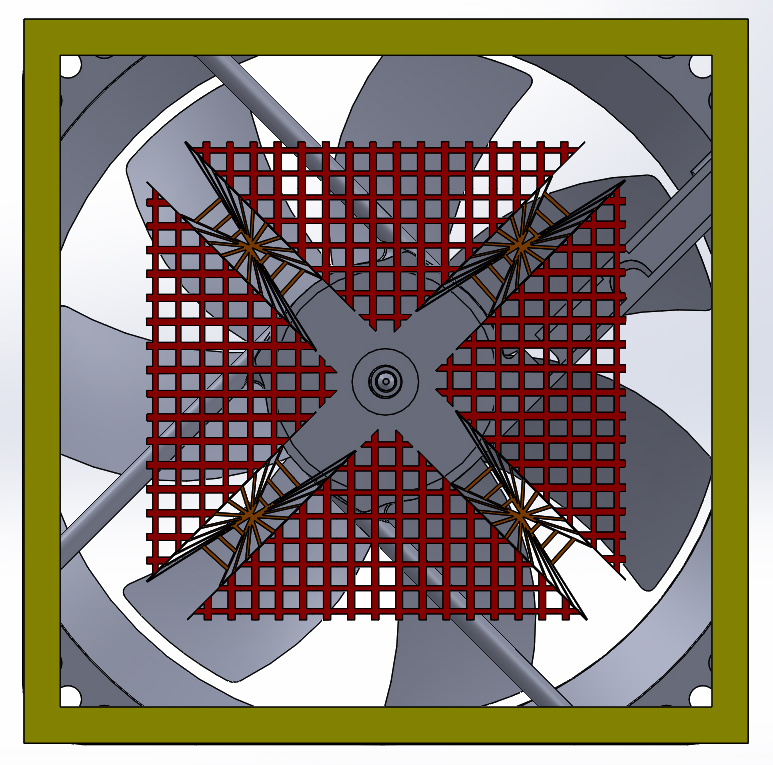}
\caption{Density based topology}
\end{figure}

\begin{figure}[h]
\centering
\includegraphics[scale=0.3]{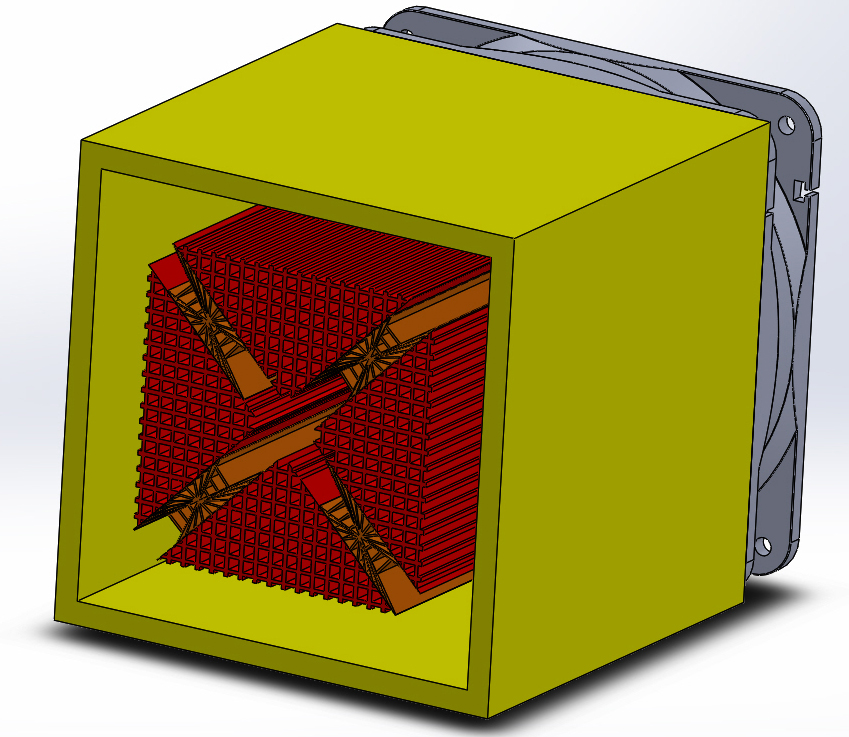}
\caption{Density based topology}
\end{figure}

\clearpage

\subsection{Further Images for parametric testing}

\begin{figure}[h]
\centering
\includegraphics[scale=0.5]{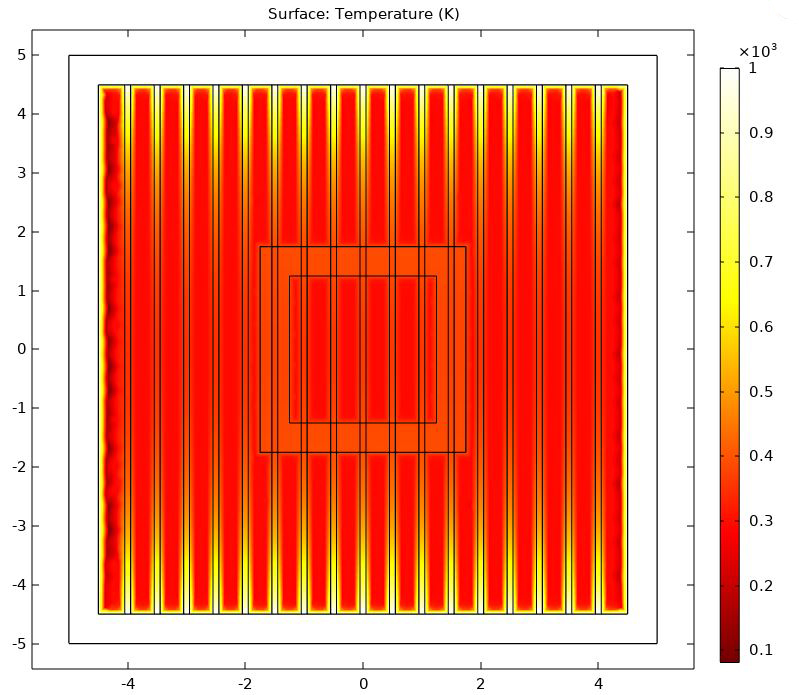}
\caption{1B Heat - Non optimised}
\end{figure}

\begin{figure}[h]
\centering
\includegraphics[scale=0.5]{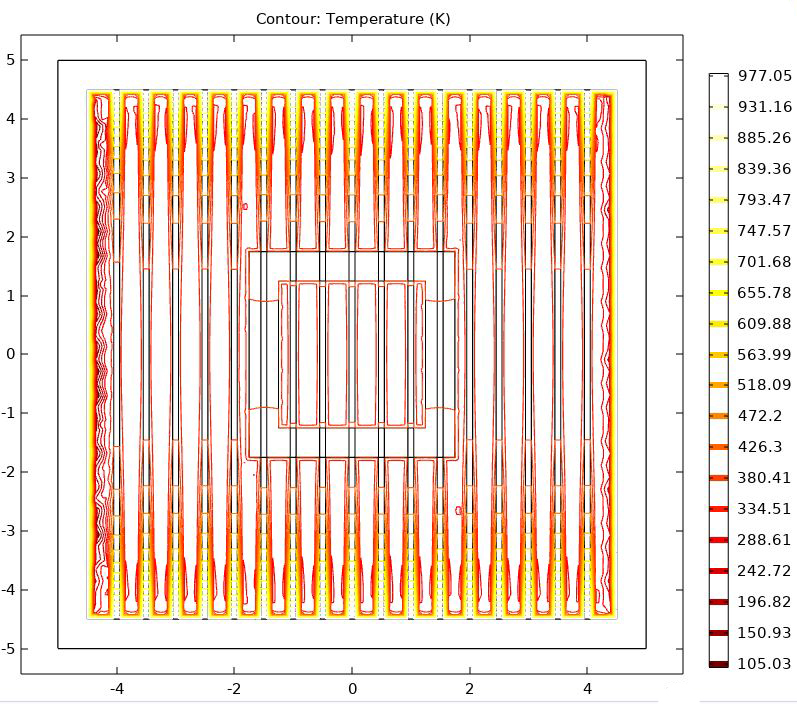}
\caption{1B Contour - Non optimised}
\end{figure}

\begin{figure}[h]
\centering
\includegraphics[scale=0.5]{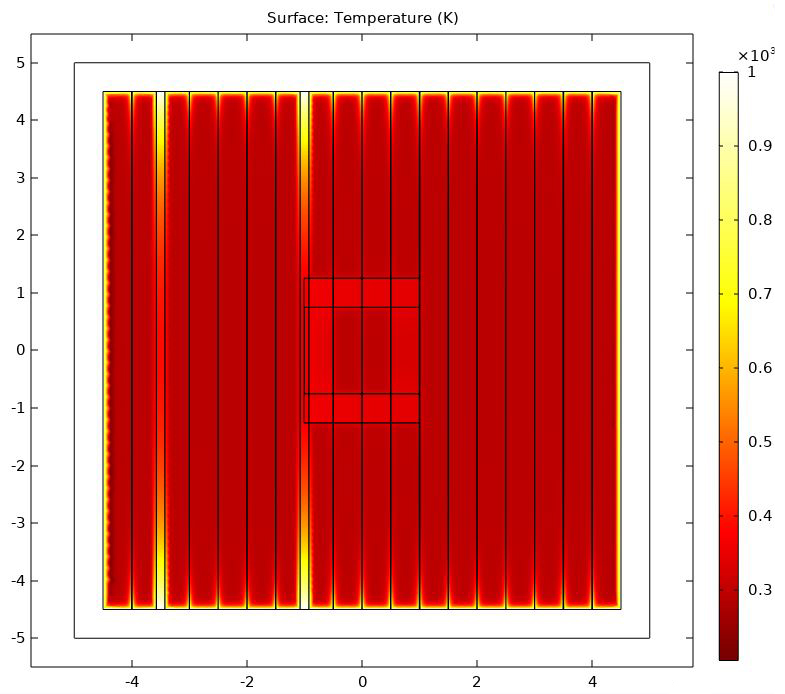}
\caption{1B Heat - optimised}
\end{figure}

\begin{figure}[h]
\centering
\includegraphics[scale=0.5]{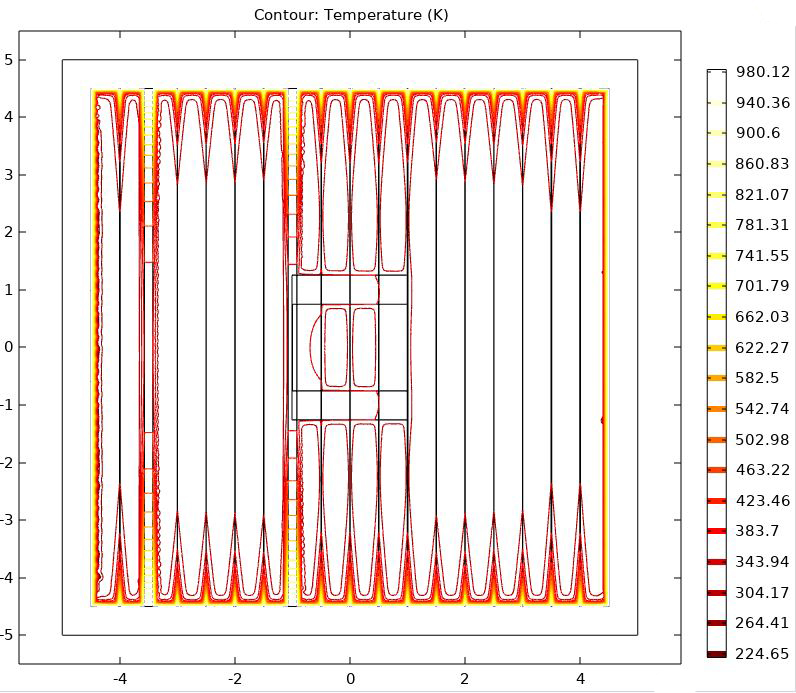}
\caption{1B Contour - optimised}
\end{figure}

\begin{figure}[h]
\centering
\includegraphics[scale=0.5]{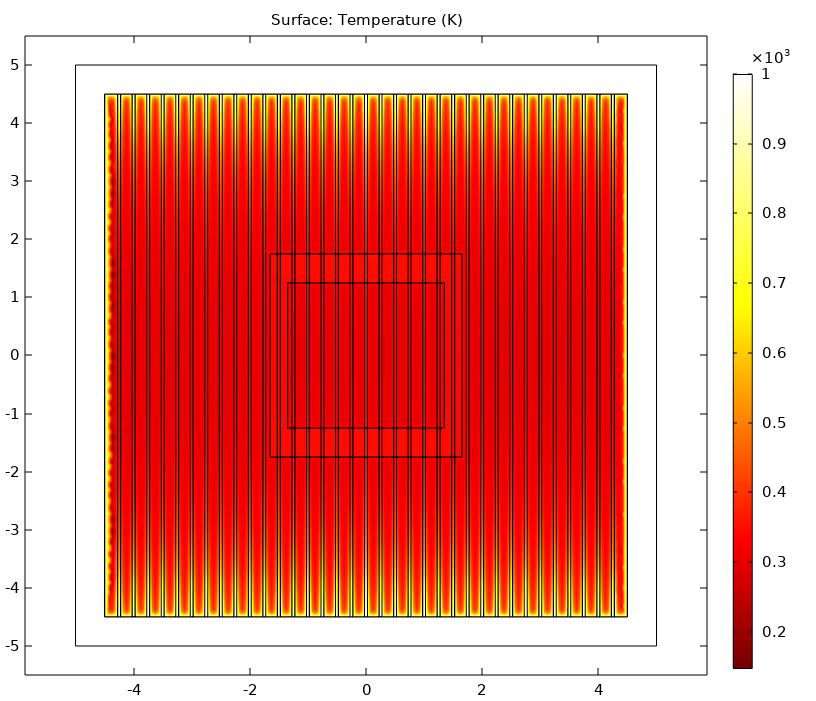}
\caption{2B Heat - Non optimised}
\end{figure}

\begin{figure}[h]
\centering
\includegraphics[scale=0.5]{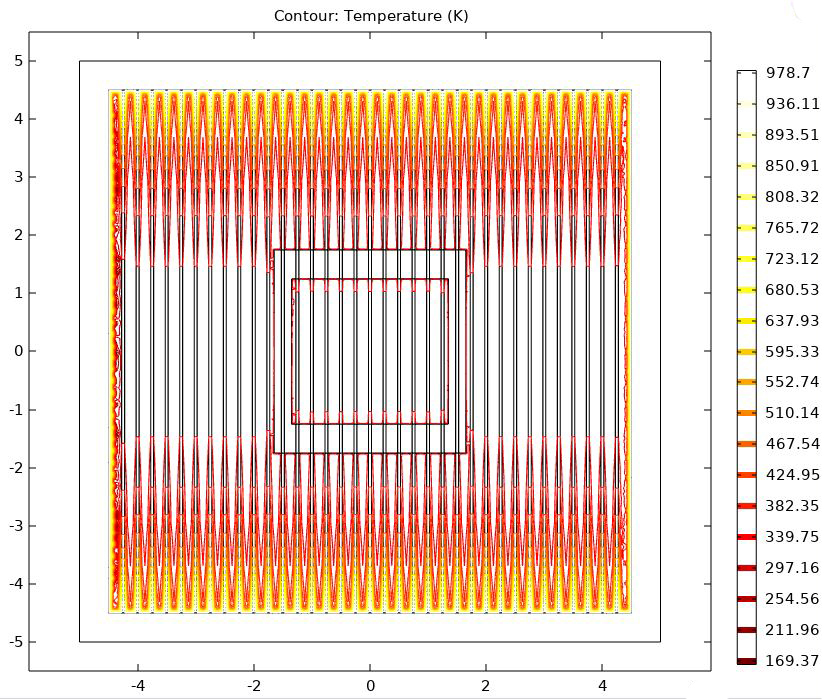}
\caption{2B Contour - Non optimised}
\end{figure}

\begin{figure}[h]
\centering
\includegraphics[scale=0.9]{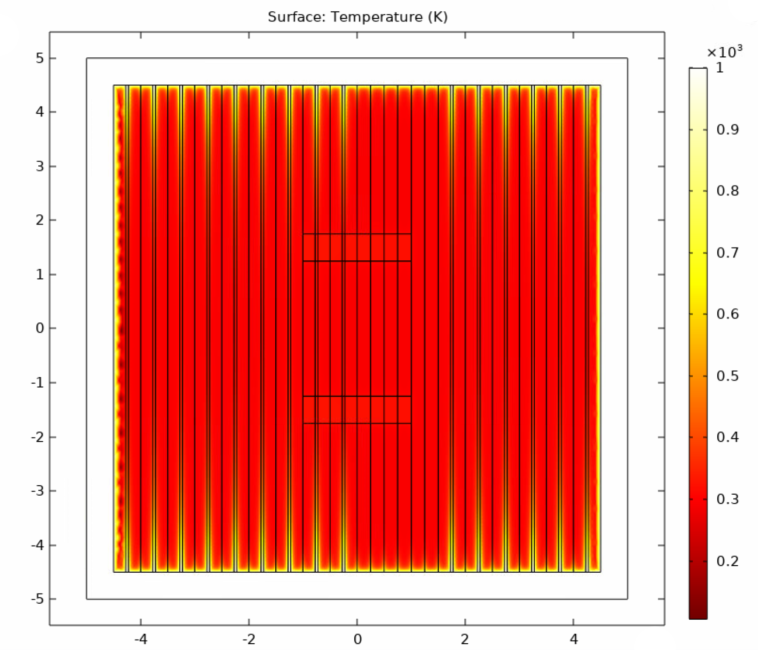}
\caption{2B Heat - optimised}
\end{figure}

\begin{figure}[h]
\centering
\includegraphics[scale=0.9]{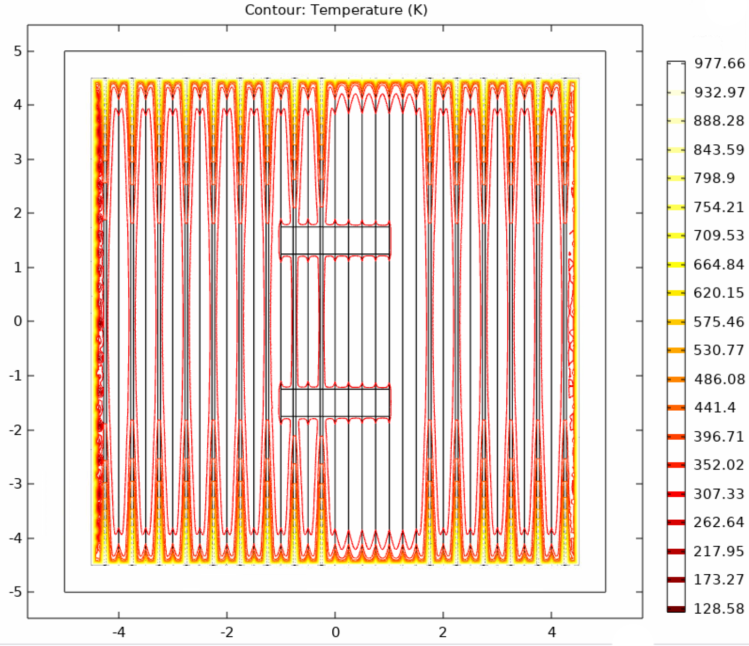}
\caption{2B Contour - optimised}
\end{figure}

\end{document}